\documentclass[journal]{IEEEtran}

\usepackage{latexsym,amsmath,amsthm,amssymb,mathrsfs}
\usepackage{amsmath, amssymb}
\usepackage{caption}
\usepackage{graphicx}
\usepackage{hyperref}
\usepackage{subcaption}
\usepackage{placeins}
\usepackage{breqn}
\usepackage{tikz}
\usepackage{algorithm}
\usepackage{algpseudocode}
\usepackage{afterpage}
\usepackage{comment}
\usepackage{soul} 
\usepackage{mathtools}
\usepackage{xargs} 
\usepackage{cite}

\usetikzlibrary{positioning}
\newcommandx{\myparagraph}[1]{\paragraph{#1}}
\newcommandx{\mysubparagraph}{\bigskip \indent}
\newcommand{\real}{\ensuremath{\mathbb{R}}}
\newcommand{\s}{\ensuremath{\mathbb{S}}}

\newcommand{\LL}{\ensuremath{\mathbb{L}}}
\newcommand{\inner}[2]{\left\langle #1,#2 \right\rangle}
\DeclareMathOperator*{\argmin}{\arg\min}

\hypersetup{colorlinks=true, linkcolor=blue, citecolor=red}
\newtheorem{theorem}{Theorem}[section]
\newtheorem{lemma}[theorem]{Lemma}

\theoremstyle{definition} 
\theoremstyle{definition} 
\theoremstyle{remark}

\AtBeginDocument{%
  \hypersetup{%
    linkcolor=[RGB]{0,150,0},%
    citecolor=blue,%
  }%
}%

\newcommand\blfootnote[1]{%
  \begingroup
  \renewcommand\thefootnote{}\footnote{#1}%
  \addtocounter{footnote}{-1}%
  \endgroup
}
\begin{document}

\title{Alignment of Continuous Brain Connectivity}

\author{Martin Cole, Yang Xiang, Will Consagra,  Anuj Srivastava, Xing Qiu and Zhengwu Zhang}

\maketitle
\IEEEpeerreviewmaketitle

\begin{abstract}
    Brain networks are typically represented by adjacency matrices, where each node corresponds to a brain region. In traditional brain network analysis, nodes are assumed to be matched across individuals, but the methods used for node matching often overlook the underlying connectivity information. This oversight can result in inaccurate node alignment, leading to inflated edge variability and reduced statistical power in downstream connectivity analyses. To overcome this challenge, we propose a novel framework for registering high-resolution continuous connectivity (ConCon), defined as a continuous function on a product manifold space—specifically, the cortical surface—capturing structural connectivity between all pairs of cortical points. Leveraging ConCon, we formulate an optimal diffeomorphism problem to align both connectivity profiles and cortical surfaces simultaneously. We introduce an efficient algorithm to solve this problem and validate our approach using data from the Human Connectome Project (HCP). Results demonstrate that our method substantially improves the accuracy and robustness of connectome-based analyses compared to existing techniques.
    \end{abstract}
    

\blfootnote{
\textit{Martin Cole and Yang Xiang contributed equally to this work. Please correspond to Dr. Zhengwu Zhang at zhengwu\_zhang@unc.edu.} \\
Martin Cole is with the Department of Psychiatry, University of Rochester, Rochester, NY 14627 USA. \\
Yang Xiang and Zhengwu Zhang are with the Department of Statistics and Operations Research, University of North Carolina at Chapel Hill, Chapel Hill, NC 27599 USA.\\
Will Consagra is with the Department of Statistics, University of South Carolina, Columbia, SC 29208 USA. \\
Anuj Srivastava is with Department of Statistics, Florida State University, Tallahassee, FL 32306 USA. \\
Xing Qiu is with the Department of Biostatistics and Computational Biology, University of Rochester, Rochester, NY 14627 USA.\\
}

    \section{Introduction}
    Image registration or alignment is a fundamental step for a variety of biomedical imaging analysis tasks \cite{toga2001role}. In this paper, we consider brain networks and their alignment. The brain network is often represented as an adjacency matrix $A \in \real^{V\times V}$, where $V$ represents the number of nodes. There are mainly two types of brain networks, functional connectivity (FC) and structural connectivity (SC), depending on how the connectivity strengths are defined in the adjacency matrix $A$. For FC, one can compute the connection strength based on the Pearson correlation between the averaged blood-oxygen-level-dependent (BOLD) signals for a pair of nodes. The BOLD signals are obtained from functional magnetic resonance imaging (fMRI). In contrast, SC is derived from diffusion MRI (dMRI), and it quantifies the degree to which brain regions are interconnected by white matter fiber bundles. Both SC and FC serve as crucial biomarkers in understanding how the brain works under various conditions \cite{sporns2005human,van2015human,elam2021human}.  
    
    Due to the varying sizes and shapes of the brains, alignment is necessary for performing group-wise analysis for a population of brain networks. The current literature largely overlooks the problem of aligning brain networks as the task becomes computationally prohibitive with a large number of nodes \cite{gold1996graduated, jiang2017graph, chen2020graph, yi2021alignment}. In contrast, most existing methodologies employ a different approach, where a predefined atlas is warped into each individual brain to pinpoint nodes or regions of interest (ROIs) \cite{de2013parcellation}. This approach assumes that the defined node matches the same brain region across individuals. 
    However, this approach suffers from two significant issues. First, the choice of the atlas is often arbitrary, and there is no universally accepted standard atlas for brain connectivity analysis \cite{messe2020parcellation, bryce2021brain}. This lack of consensus can lead to inconsistencies in analysis \cite{zalesky2010whole}. Second, the atlas warping process typically does not incorporate connectivity information, which can lead to node misalignment and inflated variances across subjects, diminishing the effectiveness of subsequent statistical analyses.
    \par
    In this work, we introduce a new approach to brain network alignment that overcomes the aforementioned challenges by using a continuous connectivity (ConCon) model, a recent representation of the brain connectivity \cite{gutman2014, moyer2017, cole2021, mansour2022}.  Let \( \mathcal{S}_1 \) and \( \mathcal{S}_2 \) be the left and right white surfaces, respectively ({as shown in Fig. \ref{fig:Figure1} Panel B), and let \( \widetilde{\Omega} = \mathcal{S}_1 \cup \mathcal{S}_2 \).
    Different from traditional ROI-based connectivity, a ConCon \( f \) defines the connection strength between any pair of points \( (x,y) \in \widetilde{\Omega} \times \widetilde{\Omega} \), i.e., \( f(x,y) \) is a continuous and symmetric function defined on \( \widetilde{\Omega} \times \widetilde{\Omega}  \). For simplicity, we study only SC and its registration and leave the extension to FC for future work. 
    
    \subsection{Background on ConCon}
    Using dMRI and structural MRI (i.e., T1-weighted MRI), tractography can be performed to infer the underlying white matter pathways that connect different brain regions \cite{fanzhang2022}. Diffusion MRI measures the random motion of water molecules in tissue, which is more restricted along the direction of white matter fiber bundles due to their organized, cable-like structure. Tractography uses this directional information to computationally reconstruct these fiber bundles as streamlines—continuous, three-dimensional curves that trace the estimated paths of white matter tracts from one brain region to another. These streamlines represent the trajectories of axonal connections within the brain’s white matter. Collectively, the resulting streamlines form the structural connectivity, providing a map of how different brain regions are physically linked.  Fig. \ref{fig:Figure1} Panel (A) shows the streamlines obtained from a random HCP subject. 
    
    In practice, to ensure the streamlines have ending points on the cortical surface $\widetilde{\Omega}$, we utilized the surface-enhanced tractography (SET) algorithm \cite{st2018surface}. Denote the pair of endpoints $(\tilde{p}_1,\tilde{p}_2)$, where $\tilde{p}_i \in \widetilde{\Omega}$, and  define the set of ending points for streamlines connecting $\widetilde{\Omega}$ as  $\tilde{O}_i=\{ (\tilde{ p}^1_1,\tilde{ p}^1_2), \cdots, (\tilde{ p}^{n_i}_{1},\tilde{ p}^{n_i}_2) \}$ for subject $i$. Here $n_i$ is the total number of streamlines connecting cortical surface $\tilde{\Omega}$. 
     Since $\mathcal{S}_1$ and $\mathcal{S}_2$ are homeomorphic to \( \mathbb{S}^2 \), we parametrize them using spherical coordinates. Let \( (p_1, p_2) \) be the image of \( (\tilde{p}_1, \tilde{p}_2) \) on \( \mathbb{S}^2_1 \cup \mathbb{S}^2_2 \) under the homomorphism, then we can denote the set of parameterized endpoints as $O_i = \{ (p^1_1, p^1_2), \ldots, (p^{n_i}_1, p^{n_i}_2) \}$,
    and $O_i \subset \Omega \times \Omega$ and $\Omega = \mathbb{S}^2_1 \cup \mathbb{S}^2_2$. 
    
    \begin{figure}[t]
        \centering
        \includegraphics[width=0.5\textwidth]{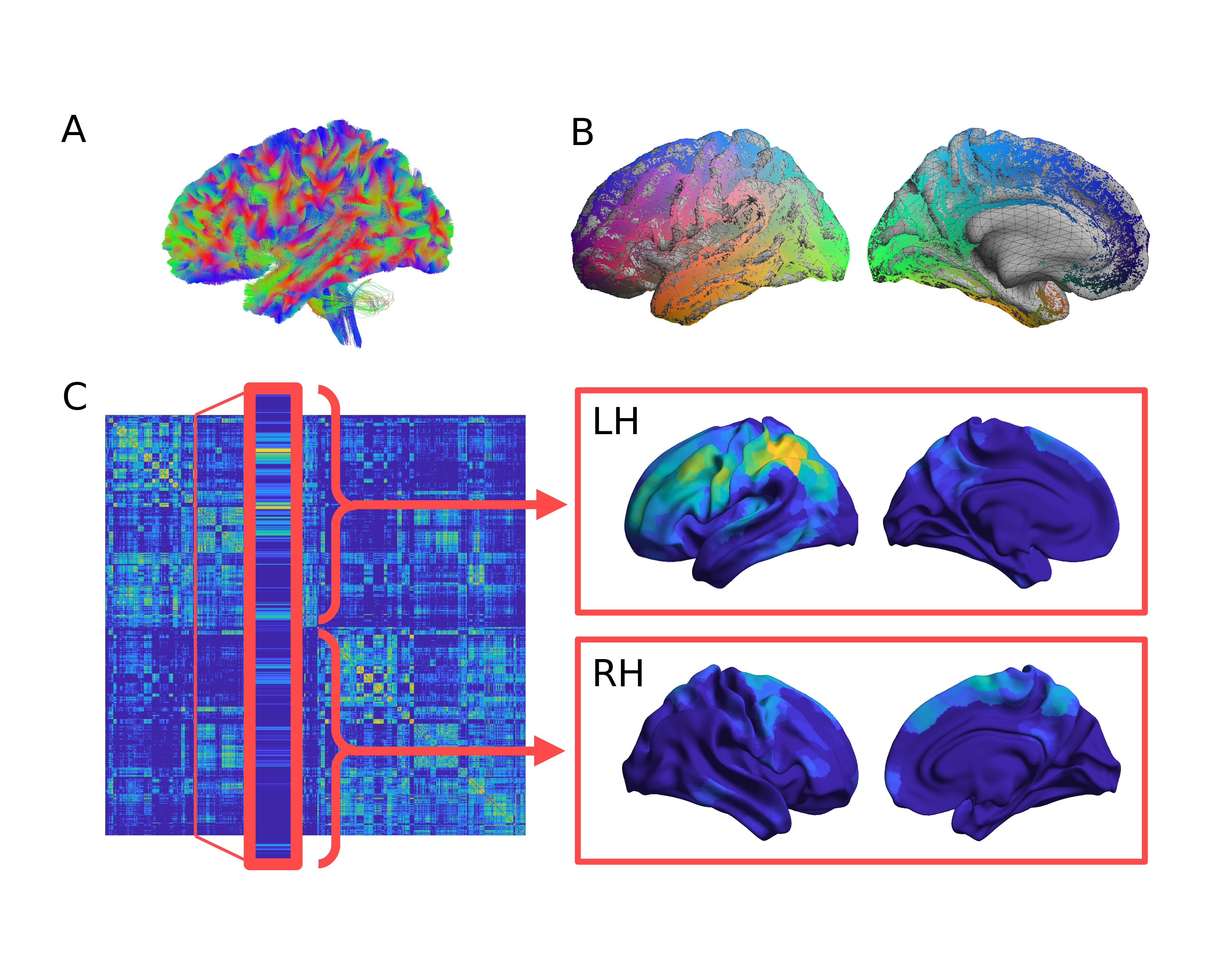}
        \caption{\small{Panel A shows the tractography of a random subject, panel (B) shows where the tractography intersects with the cortical surface, and panel (C) shows the ConCon matrix of a random subject and the projection of a single column onto the cortical surface.}}
        \label{fig:Figure1}
    \end{figure}
    
    
    The endpoints set $O$ completely describes the surface connectivity pattern generated by the estimated streamlines from SET. In the ConCon framework, the endpoints \( O \) are modeled as a realization of a stochastic point process on \( \Omega \times \Omega \) with an unknown but integrable intensity function \( f^{*}: \Omega \times \Omega \rightarrow [0, \infty) \) \cite{moyer2017continuous, cole2021, mansour2022, consagra2022analyzing}. This function is defined as follows: for any two measurable regions \( E_1 \subset \Omega \) and \( E_2 \subset \Omega \), denote \( N(E_1, E_2) \) as the number of streamlines ending in \( (E_1, E_2) \), then \( f^{*} \) satisfies the following equation:
    \begin{equation} \label{eqn:first_moment_assumption}
    \mathbb{E}\left[N(E_1,E_2)\right] = \int_{E_1} \int_{E_2} f^{*}(x,y) \, dx \, dy < \infty.
    \end{equation} 
    Associated with intensity function $f^{*}$ is density function $f$ defined by:
    $$
    f = \frac{f^{*}(x,y)}{\int_{\Omega}\int_{\Omega} f^{*}(u,v) \, du \, dv},
    $$ 
    which normalizes $f^*$. This latent function \( f \) is referred to as ConCon and is the object of interest in this work.
    
    Crucially, the ConCon representation is \textit{resolution agnostic} and does not rely on a pre-defined brain atlas and so avoids the previously mentioned issues resulting from ROI-based connectivity. Moreover, it has been demonstrated that ConCon is a more powerful representation for a variety of population-based neuroscience analyses, including trait prediction and group-wise hypothesis testing \cite{consagra2023continuous}.
    
    \subsection{Alignment of ConCon Functions}
    
    Let $\{ f_1, \dots, f_N \}$ be the set of observed ConCon functions derived from \( N \) subjects on the individual white surfaces \( \widetilde{\Omega}_1, \ldots, \widetilde{\Omega}_N \). Given that these white surfaces exhibit varying shapes and folding patterns, it becomes essential to determine the correspondence between \( f_i(x, y) \) for \( x, y \in \widetilde{\Omega}_i \) and \( f_j(x', y') \) for \( x', y' \in \widetilde{\Omega}_j \) in order to compare \( f_i \) and \( f_j \). 
    
    We now discuss two possible ways of aligning ConCon functions $f_1$ and $f_2$  on $\widetilde{\Omega}_1 \times \widetilde{\Omega}_1$ and $\widetilde{\Omega}_2 \times \widetilde{\Omega}_2$, respectively. The first way is to deform the domains to a template surface $\widetilde{\Omega}_{\mu}$ and map any data on the individual domain to the template domain $\widetilde{\Omega}_{\mu}$. This idea has been previously used in \cite{lila2022functional} for jointly modeling brain shape and functional connectivity. However, the deformation of $\widetilde{\Omega}_i$ to a template considers only its shape information. Without considering the connectivity features encoded in $f_i$, the method only provides a suboptimal alignment of the connectivity.
    
    The second way is to keep the shape of $\widetilde{\Omega}_i$ untouched but change its parameterization to align the ConCon functions. More specifically, since the white surfaces are homeomorphic 2-spheres, we can parameterize points on $\mathcal{S}_j$ using spherical coordinates, and thus $f_i(x,y)$ becomes functions defined on $\Omega \times \Omega$ for $\Omega = \s^2_1 \cup \s^2_2$. This parameterization process establishes an initial correspondence between $f_1$ and $f_2$, i.e., $f_1(x,y)$ is matched with $f_2(x,y)$ for $x,y \in \Omega$ . The alignment of $f_2$ to $f_1$ is now to find an optimal reparameterization or warping function $\gamma_\Omega$ such the warped $f_2$ denoted as $(f_2*\gamma_\Omega)$ (whose precise definition will be given later) is best matched with $f_1$, where $\gamma_\Omega$ is a diffeomorphism of $\Omega$.
    
    Formulating the brain connectivity alignment using the latter method has many advantages over the other way: 1) It directly aligns the connectivity data, and no other features are involved in confounding the alignment process; 2) the original surface $\tilde{\Omega}_i$ is unchanged, and we can easily and accurately map any statistical findings back to the original surface to interpret the findings; and 3) it formulates the brain connectivity alignment problem into a functional data alignment problem, enabling us to borrow ideas from existing solutions to solve this challenging problem \cite{srivastava2016functional}.

    \subsection{Our Approach}
    We work on the second method to align brain networks. 
    Once the brain networks are represented using ConCon functions, existing methods from functional data alignment can be adapted to our situation.  For example, \cite{liu2004functional} form the functional alignment problem on $[0,1]$ as finding $\Gamma_{[0,1]}(t)$'s in the form of ${\int_0^t |g_i^{(v)}(s)|^pds}/{\int_0^1|g_i^{(v)}(s)|^pds}$ (with $v = 0$ and $p=1$ as recommended values) so that $g_i\circ \Gamma_{[0,1]}$'s are aligned, where $g_i: [0,1] \mapsto \real$. \cite{JamesAOAS127} uses moment-based matching for aligning 1D functions on $[0,1]$. 
    \cite{srivastava2010shape} and \cite{joshi2007novel} use a square-root velocity function (SRVF) to align and analyze the shapes of curves, i.e., $g: [0,1] \mapsto \real^n$. More recently,  the Fisher-Rao Riemannian metric has been used to separate the phase and amplitude parts of 1D functional data on $[0,1]$ in \cite{srivastava2011registration}. Alignment of functional data on the manifold, i.e., $g:[0,1]\mapsto \mathcal{M}$ where $\mathcal{M}$ is a nonlinear manifold, has been investigated in \cite{zhang2018rate,zhang2018phase,zhang2022amplitude}. However, the ConCon function $f$ is defined on a product manifold domain $\Omega \times \Omega$, which is significantly different from the previous works. 
    
For simplicity, we utilize a much simpler domain $\Omega = [0,1]$ to discuss the problem and derive all theoretical and computational tools for solving the alignment problem on $[0,1] \times [0,1]$. We then extend the solution to domains $\s^2$ and $\s^2_1 \cup \s^2_2$.   We refer to our method as mEtric-based coNtinuous COnnectivity REgistration (ENCORE). 
Our main contributions can be summarized as follows: 
\begin{enumerate} \item To the best of our knowledge, this is the first study to formulate brain network registration as a problem of aligning functional data on a product manifold. While \cite{gutman2014} represents an early exploration in this direction, our approach is significantly more comprehensive in scope and methodology.
\item We develop an efficient and fast gradient-based algorithm in ENCORE to optimize the alignment of cortical surfaces based on connectivity patterns. This alignment facilitates improved brain network registration, enabling more robust statistical analyses.
\item Through extensive simulations and real-world data analyses, we demonstrate the effectiveness of ENCORE in aligning brain networks.
\end{enumerate}

The remainder of the paper is organized as follows. In Section \ref{sec:algI}, we introduce ConCon alignment on \(\Omega = [0,1]\) and extend the method to \(\Omega = \mathbb{S}^2\) in Section \ref{sec:s2extension}, followed by an extension to \(\Omega = \mathbb{S}^2_1 \cup \mathbb{S}^2_2\) in Section \ref{sec:extens2us2}. In Section \ref{sec:deftemp}, we present a method for defining a template from a set of ConCon functions. We then provide extensive simulation studies in Section \ref{sec:simulation} and real-data analyses in Section \ref{sec:realdata}. Finally, we conclude the paper in Section \ref{sec:conclusiondis}.

\section{ConCon Alignment on $\Omega = [0,1]$}
\label{sec:algI}
We start with $\Omega = [0,1]$ and define the following notations to help formulate the research problem. \begin{itemize}
    \item $\Gamma_{[0,1]}$ - the set of boundary-preserving diffeomorphisms on $[0,1]$. A function $\gamma \in \Gamma_{[0,1]}$ satisfies the following conditions: it is a mapping from $[0,1]$ to $[0,1]$ with $\gamma(0)= 0$ and $\gamma(1)= 1$, and $\gamma$ is invertible and both $\gamma$ and its inverse are smooth.  The elements of $\Gamma_{[0,1]}$ form a group, i.e., 1) for any $\gamma_1, \gamma_2 \in \Gamma_{[0,1]}$, their composition $\gamma_1 \circ \gamma_2 \in \Gamma_{[0,1]}$; 2) for any $\gamma \in \Gamma_{[0,1]}$, its inverse $\gamma^{-1} \in \Gamma_{[0,1]}$; and 3) the identity function $\gamma_{id}(x) =x $ is an element in $\Gamma_{[0,1]}$. 
    
    \item $\mathcal{F}_{I}$ - the set of ConCon functions on $[0,1] \times [0,1]$. For  $f$ in $\mathcal{F}_{I}$, it has to satisfy several constraints: 1) the symmetry constraint: $f(x,y) = f(y,x)$; 2) the non-negativity constraint: $f(x,y) \geq 0$; and 3) the constant integration constraint: $\int_{\Omega} \int_{\Omega} f(x,y) dxdy = C$, where $C$ denotes a constant. The condition 1) implies that the connectivity is undirectional, 2) implies that the connection strength between any pair of nodes is always non-negative, and 3) implies that we have normalized the ConCons to remove the scaling confounding. In this paper, we let $C = 1$, and the ConCon function is a probability density function (PDF), and for two disjoint regions $E_1,E_2 \subset \Omega$, the integration $P(E_1,E_2) = \int_{E_1}\int_{E_2} f(x,y) dxdy$ describes the connection strength between $E_1$ and $E_2$. 

    \item  $f \circ \tilde{\gamma} (x, y) \equiv f(\gamma(x),\gamma(y))$ - a compact notation for applying an element $\gamma \in \Gamma_{[0,1]}$ to warp ConCon functions.   Elements of $\Gamma_{[0,1]}$ define the following mapping on this domain:
   $ \tilde{\gamma} : [0,1] \times [0,1] \rightarrow [0,1] \times [0,1], 
    (x, y) \mapsto (\gamma(x), \gamma(y)).$ Therefore a compact notation can be obtained  $f \circ \tilde{\gamma} (x, y) \equiv f(\gamma(x),\gamma(y))$. 
\end{itemize}

Considering the constraints associated with ConCons, we propose the following natural group action of $\Gamma_{[0,1]}$ on $\mathcal{F}_{I}$:

    $$ (f*{\gamma}) = (f\circ \tilde{\gamma})\dot{\gamma}(x)\dot{\gamma}(y).$$
    Under this action, we can easily show that all constraints of $f \in \mathcal{F}_I$ get preserved. 
The value $f(x,y)$ represents the probability density of the streamline connections between $x$ and $y$. A warping $\gamma$ will stretch or squeeze the underlying domain, thereby altering the corresponding density.  Fig. \ref{fig:warp} panel (A) shows an $f \in \mathcal{F}_I$, panel (B) shows a simulated $\gamma \in \Gamma$ and panel (C) shows the warped function  $(f*\gamma)$. }

\begin{figure}[t]
    \centering
    \includegraphics[width=0.5\textwidth]{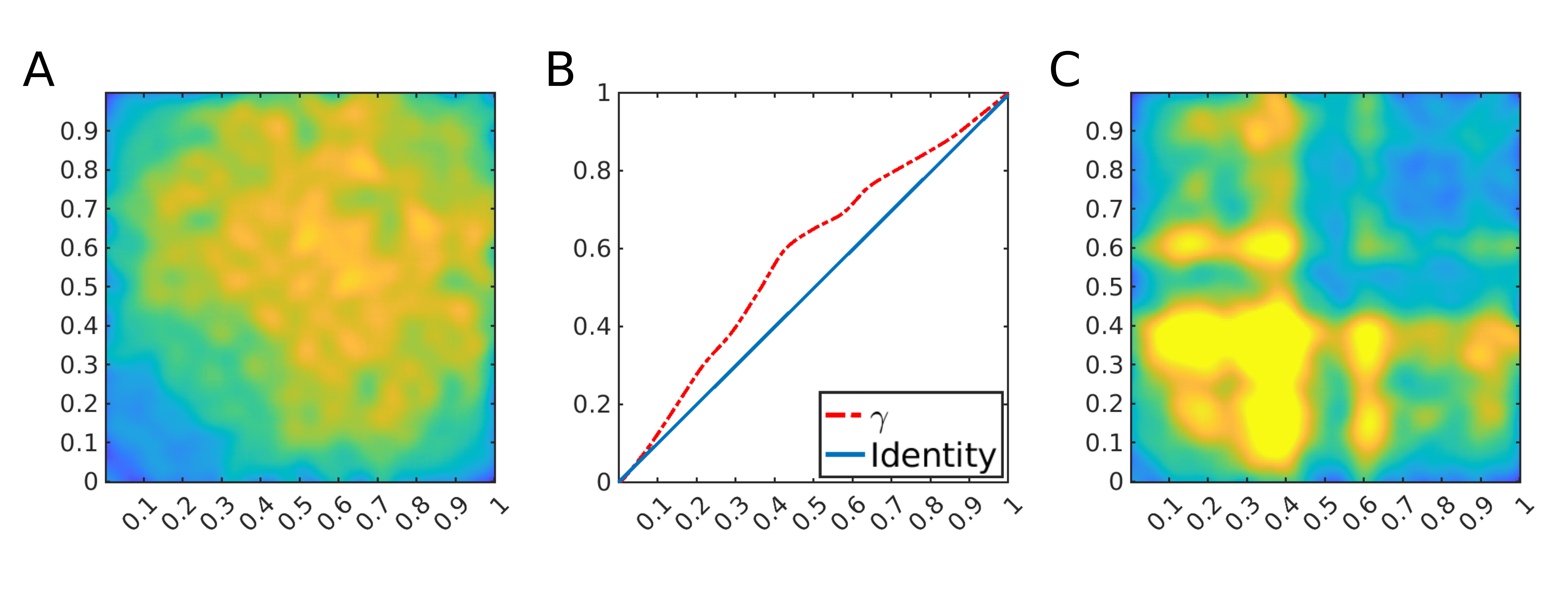}
    \caption{\small{Panel A shows a simulated ConCon function $f$, panel B shows a randomly generated warping function $\gamma$ for $\Omega = [0,1]$, and panel (c) shows the warped ConCon $(f*\gamma)$.}}
    \label{fig:warp}
\end{figure}

Now, let us formulate the registration problem. With the warping action defined, the registration between $f_1$ and $f_2$ can be formulated as a problem to find a $\gamma \in \Gamma_{[0,1]}$ such that $(f_2*\gamma)$ is optimally matched with $f_1$. To solve it, we face two important problems: a) {\it how to quantify the matching between $(f_2*\gamma)$ and $f_1$}, and b) {\it how to find a $\gamma^* \in \Gamma_{[0,1]}$ to optimize the criterion defined in} a). 

The problem a) relates to finding an appropriate metric to quantify the difference between $f_2$ and $f_1$. For example, the simplest way to compute the distance between two functions is to use the standard $\mathbb{L}^2$ distance for functions, i.e., $\|f_1 - f_2\| = (\iint(f_1 - f_2)^2 dx dy)^{1/2}$. However, it is easy to show that in general, the action of $\gamma$ does not preserve the $\mathbb{L}^2$ distance in $\mathcal{F}_I$: using the same warping function $\gamma$, $\left\Vert f_1(x,y) - f_2(x,y) \right\Vert \neq \left\Vert (f_1*{\gamma}) -(f_2*{\gamma}) \right\Vert$. This lack of isometry is problematic because we wish the distance between two ConCon functions is kept invariant before and after the registration (warping). Since $f_i$'s are PDFs, another potential choice is the Kullback-Leibler (KL) divergence. However, the KL divergence is not a proper distance because it is not symmetric, meaning that $\mathrm{KL}(f_1, f_2) \ne \mathrm{KL}(f_2, f_1)$ , which is problematic in the context of registration.

It turns out that if we apply simple square-root transformation to functions in $\mathcal{F}_I$, the warping action will preserve the desired isometry property. This fact will be made evident in the following sections.

\subsection{Square-Root Representation of ConCon}
\label{sec:srvf_concon}

Instead of directly working with the function $f$, we define a square-root transformation of $f$ with the mapping:
\begin{align}
\label{eqn:Qtrans}
    &Q : \mathbb{R} \rightarrow \mathbb{R}, \;\;
    Q(s) = \begin{cases}
                \sqrt{s}, \;\; &\text{if } s \ge 0\\
                -\sqrt{-s}, \;\; &\text{otherwise}.
            \end{cases}
\end{align}

\noindent Note that for functions in $\mathcal{F}_{I}$ there is no negative values.  Applying the $Q$ mapping, our new function space is defined as
\begin{equation}\label{eq:qmap}
    \mathcal{Q}_I \equiv \left\{q(x,y) = Q\left(f(x,y)\right) \; | \; f\in \mathcal{F}_{I}\right\}.
\end{equation}
Since $f$ integrates to one, we have $\iint q(x,y)^2 dxdy = \iint f(x,y) dxdy = 1$. That is, the $q$ function is an element of the positive orthant $\s^\infty$. The $q$ function is also referred to as the {\it half density} of $f$, and from the half density, we can easily obtain the full density using $f(x,y) = q(x,y)^2$.  Since $\s^{\infty}$ is a unit sphere, it has a very nice geometry. With the $\mathbb{L}^2$ Riemannian metric, the distance between two elements $g_1,g_2$ on $\s^{\infty}$ can be quantified as $$d_{R}(g_1,g_2) = \cos^{-1}(\iint g_1g_2 dxdy).$$ Since $\mathcal{Q}_I \subset \s^{\infty}$, we use the same metric to quantify the distance between two functions $q_1$ and $q_2$ in $\mathcal{Q}_I$: $d_{R}(q_1,q_2) = \cos^{-1}(\iint q_1q_2 dxdy)$. 

\subsection{Warping Invariant Riemannian Distance between $Q$-transformed ConCons}
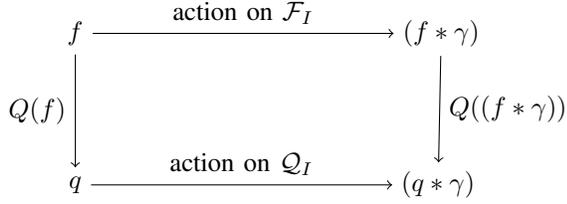
\begin{figure}
    \centering
    \begin{tikzpicture}[node distance={1.5cm and 4cm}, auto]
        \node (R2a) {$q$};
        \node (R2b) [right=of R2a] {$(q * \gamma)$};
        \node (R3a) [above=of R2a] {$f$};
        \node (R3b) [right=of R3a] {$(f * \gamma)$};

        \draw[->] (R2a) to node {action on  $\mathcal{Q}_I$} (R2b);
        \draw[->] (R3a) to node [midway,left]{$Q(f)$} (R2a);
        \draw[->] (R3a) to node {action  on $\mathcal{F}_I$} (R3b);
        \draw[->] (R3b) to node {$Q((f * \gamma))$} (R2b);
    \end{tikzpicture}
    \caption{{\small A commutative diagram to represent the relationship between functions $f \in \mathcal{F}_{I}$, functions $q \in \mathcal{Q}_I$, and their respective group actions by an element of $\gamma$ from $\Gamma_{[0,1]}$.}}
  \label{fig:QCgamma}
\end{figure}

If a ConCon function is warped (or reparameterized), e.g., $(f*{\gamma})$, then the corresponding Q-function  can be derived as: $(q*{\gamma}) = Q((f*\tilde{\gamma})) = (q\circ \tilde{\gamma})\sqrt{\dot{\gamma}(x)}\sqrt{\dot{\gamma}(y)}$. In fact, this denotes an action of group $\Gamma_{[0,1]}$ on $\mathcal{Q}_I$ from the right side: $\mathcal{Q}_I \times \Gamma_{[0,1]} \mapsto \mathcal{Q}_I$ by $(q*{\gamma}) = (q\circ \tilde{\gamma})\sqrt{\dot{\gamma}(x)}\sqrt{\dot{\gamma}(y)}$. Figure \ref{fig:QCgamma} illustrates two different possible paths to get a warped $q$ function. We then can derive that under the Q-transformation, the action of warping under the $\LL^2$ Riemannian metric is an isometry. 

\begin{lemma}
    For any two ConCon functions $f_1$, $f_2$, and their corresponding Q-transformed functions $q_1$ and $q_2$, with the simple  $\LL^2$  Riemannian metric, we have $d_{R}(q_1,q_2) = d_{R}((q_1*{\gamma}),(q_2*{\gamma}))$.
    \end{lemma}
    
    See the proof in Supplementary S1. This isometry property forms an important foundation for our ConCon registration problem. We can just use $d_{R}(\cdot,\cdot)$ as the energy function to quantify the matching between $f_1$ and $f_2$, e.g., denote $H(f_1,f_2) = d_{R}(q_1,q_2)$. To align $f_2$ to $f_1$, we simply fix $f_1$ and find the optimal $\gamma^*$ to warp $f_2$ to minimize the energy function, i.e., $$ \gamma^* = \arg \min_{\gamma \in \Gamma_{[0,1]}} H(f_1,(f_2*\gamma)).$$ We can also show that if $\gamma^*$ is the optimal warping to align $f_2$ to $f_1$, then $(\gamma^*)^{-1}$ is the optimal warping to align $f_1$ to $f_2$. That is, the solution is {\it inverse consistent}. To further simplify the optimization problem, we consider the ambient space of $\mathcal{Q}_I$, the symmetric functions on $\Omega \times \Omega$, and use the standard $\LL^2$ distance in the ambient space to define the optimization problem: 
    \begin{equation} \label{eqn: warpenergy}
        \gamma^* = \argmin_{\gamma \in \Gamma_{[0,1]}} H(f_1,(f_2*\gamma)); H(f_1,(f_2*\gamma)) = \|q_1 - (q_2*{\gamma})\|^2.  
    \end{equation}
    It is easy to show that the two optimization problems are equivalent since one minimizes the arc length, and the other minimizes the chord length.
    

    \subsection{Deriving the Gradient of $H$}
    For $\Omega = [0,1]$, ENCORE's optimization problem can be easily solved using dynamic programming (DP) \cite{srivastava2010shape}. However, when the domain becomes more complex, e.g., $\Omega = \s^2$,  the DP algorithm no longer works. Here we propose a gradient-based approach that can be easily extended to more complex domains.
    
    We first derive the gradient of our cost function $H$ in (\ref{eqn: warpenergy}). Let $[q] = \{(q * \gamma) \text{, for } \gamma \in \Gamma_{[0,1]}\}$ denote the orbit of $q \in \mathcal{F}_I$ under all warping actions. We define the function 
    \begin{equation}
    \label{def:phi}
    \Phi(\gamma)(x,y) = \left(q \circ \tilde{\gamma}(x,y)\right)\sqrt{\dot{\gamma}(x)} \sqrt{\dot{\gamma}(y)}
    \end{equation}
    as a mapping from $\Gamma_{[0,1]}$ to $[q]$. Let $T_{\gamma_{id}}(\Gamma_{[0,1]})$ be the tangent space of the space of diffeomorphisms $\Gamma_{[0,1]}$ at $\gamma_{id}$ (the identity warp). Specifically, for the case where $\Omega = [0,1]$, we have $T_{\gamma_{id}}(\Gamma_{[0,1]}) = \{b \, : \, [0,1] \rightarrow \mathbb{R} \; | \; b(0) = 0,\; b(1) = 0,\; $b$ \text{ is a smooth function}\}$. Assuming that we have an orthonormal basis for $T_{\gamma_{id}}(\Gamma)$, denoted as $\{{b}_1, {b}_2, \dots\}$, we can derive the gradient of the cost function $H$ as
    \begin{equation}\label{dcost}
        \nabla H = \sum_{i=1}^{\infty} (\nabla_{b_i}H)b_i \approx \sum_{i=1}^{M} (\nabla_{b_i}H)b_i,
    \end{equation}
     for some large $M$, where $\nabla_{{b}_i}H$ represents the directional derivative of $H$ along the direction ${b}_i \in T_{\gamma_{id}}(\Gamma_{[0,1]})$. To derive $\nabla_{{b}_i}H$, we start with the derivative of $\Phi$ in Eq. (\ref{def:phi}).
    
    \begin{lemma}\label{dphi}
    The differential map of $\Phi \, : \, \Gamma_{[0,1]} \rightarrow [q]$, evaluated at $\gamma_{id}$, is a linear transformation $d\Phi_{\gamma_{id}} \, : \, T_{\gamma_{id}}(\Gamma_{[0,1]}) \rightarrow T_{q}([q])$, and is defined by the equation
    \begin{equation}
        \begin{aligned}
        (d\Phi_{\gamma_{id}}(b))(x,y) &= \begin{bmatrix}\frac{\partial q(x,y)}{\partial x} & \frac{\partial q(x,y)}{\partial y}\end{bmatrix}{\begin{bmatrix} b(x) \\ b(y) \end{bmatrix}} \\ &\quad\quad\quad + 
        \frac{q(x,y)}{2} \left(\dot{b}(x) + \dot{b}(y)\right),
        \end{aligned}
    \end{equation}
    where ${b} \in T_{\gamma_{id}}(\Gamma_{[0,1]})$, and $\dot{b}$ is its first derivative.
    \end{lemma}
    
    The proof can be found in Supplementary S2. We can now specify the tangent space $T_{q}([q])$ which can be identified through Lemma \ref{dphi}, i.e., the set $\{d\Phi_{\gamma_{id}}({b_1}), d\Phi_{\gamma_{id}}({b_2}),\dots \; | \; {b}_i \in T_{\gamma_{id}}(\Gamma_{[0,1]})\}$. The elements of this set span the space $T_{q}([q])$, forming a basis, but not necessarily an orthonormal one since the mapping $d\Phi_{\gamma_{id}}$ is not an isometry. 
    
    \begin{theorem}\label{th:dcost}
    The directional derivative of $H$ in the direction ${b}_i \in T_{\gamma_{id}}(\Gamma_{[0,1]})$ is given by 
    \begin{equation*}
        \nabla_{{b}_i} H = -2 \inner{q_1({x},{y}) - q_2(x,y)}{d\Phi_{\gamma_{id}}(b_i)},
    \end{equation*}
    where $d\Phi_{\gamma_{id}}(b_i)$ is defined through Lemma \ref{dphi}. 
    \end{theorem}
    
    The proof of Theorem \ref{th:dcost} is in Supplementary S3. 
    
    \subsection{Alignment Through Composition of Simple Warpings}
    With these results in hand, we now circle back to our primary objective of minimizing the function \( H \) utilizing a gradient descent algorithm. At each iteration, our goal is to determine an incremental \( \gamma \) within \( \Gamma_{[0,1]} \)  to bring down $H$ according to its gradient. That is, in each step, we have the differential of $H$, approximated by $\nabla H = \sum_{i=1}^{M} (\nabla_{b_i}H)b_i$ and evaluated at $\gamma_{id}$ through Theorem \ref{th:dcost},  giving us an element in $T_{\gamma_{id}}(\Gamma_{[0,1]})$. With $\nabla H$, we define the incremental warp as $\exp_{\gamma_{id}}({\sigma \nabla H})$ with a small step size $\sigma$, where $\exp_{\gamma_{id}}$ is the exponential map on $\Gamma_{[0,1]}$ evaluated at $\gamma_{id}$. For $\Omega = [0,1]$, $\exp_{\gamma_{id}}({\sigma \nabla H}) = \gamma_{id}+{\sigma \nabla H}$. Denoting the incremental group element by $\gamma_k$ for the $k$-th iteration, and cumulative group element by $\gamma^{(k)}$, we have $\gamma^{(k)} = \gamma^{(k-1)}\circ\gamma_k$.  Algorithm \ref{alg:increamentalgamma}  presents the algorithm for this method.

    \begin{algorithm}
    \caption{Align ConCons with $\Omega = [0,1]] $}\label{alg:increamentalgamma}
    \begin{algorithmic}
    \State \textbf{Input}: $f_1, f_2$ and their Q-transformations $q_1$ and $q_2$
    \State Set $k = 1$, $\gamma^{(1)} = \gamma_{id}$, and a small positive constant $\epsilon$ and step size $\sigma$.
    \vspace{0.2em}
    \While{$||\nabla H|| > \epsilon$}
    \State ${q}^\prime_2 \gets (q_2 * \gamma^{(k)})$
    \vspace{0.2em}
    \State \text{Compute $\nabla H$ using Eq. \eqref{dcost} with $q_1$ and ${q}^\prime_2$}
    \vspace{0.2em}
    \State \text{Compute the incremental warp } ${\gamma}_k \gets \exp_{\gamma_{id}}({\sigma \nabla H})$ 
    \vspace{0.2em}
    \State \text{Compute overall warp } $\gamma^{(k + 1)} \gets \gamma^{(k)} \circ {\gamma}_k$
    \State $k \gets k + 1$
    \vspace{0.2em}
    \EndWhile
    \end{algorithmic}
    \vspace{0.2em}
    \textit{\noindent Here $\exp_{\gamma_{id}}$ is the exponential map on $\Gamma_{[0,1]}$ evaluated at $\gamma_{id}$, i.e., $\exp_{\gamma_{id}}({\sigma \nabla H}) = \gamma_{id}+{\sigma \nabla H}$. }
    \end{algorithm}
    The proposed Algorithm \ref{alg:increamentalgamma} breaks a complex alignment problem into a series of simpler subproblems, making the overall optimization process more manageable. Instead of computing a single, potentially complicated warping function to align \( f_2 \) to \( f_1 \), we iteratively update the transformation using incremental warping functions \( \gamma_k \). Each incremental warp is determined by the gradient descent step, ensuring a gradual and controlled deformation. The final warping function \( \gamma^* = \gamma^{(k)} \) is obtained through successive compositions of these incremental warps, ultimately leading to the desired alignment. The step size parameter \( \sigma \) (also referred to as the learning rate in the literature) controls the magnitude of each update, balancing convergence speed and stability.

    \section{ConCon alignment on $\Omega = \s^2$}\label{sec:s2extension}
    Considering applications for the human brain, we need to adapt Algorithm \ref{alg:increamentalgamma} for the domain $\Omega = \mathbb{S}^2$.
    While the core idea remains similar, we face several theoretical and computational challenges to extend the algorithm from the simple 1D interval to a non-linear manifold domain. To facilitate this extension, we first revisit the foundational concepts. Subsequently, we will delve into the computational methodologies required to run Algorithm \ref{alg:increamentalgamma} to align ConCon functions on the product space $\Omega \times \Omega = \mathbb{S}^2 \times \mathbb{S}^2$.

    Let $\gamma \in \Gamma_{\s^2}$ be a diffeomorphism from $\s^2$ to $\s^2$, where $\Gamma_{\s^2}$ is the set of all such diffeomorphisms. Let $\mathcal{F}_{\s^2}$ be the set of all PDFs defined on $\s^2 \times \s^2$. Fig. \ref{fig:vector_field_comparison} shows two examples of $\gamma$ as vector fields. 
    The warping action $(f*\gamma)$ for an $f \in \mathcal{F}$ is now defined as $$(f*\gamma) = (f\circ\tilde{\gamma}) {|J_\gamma(x)|}{|J_\gamma(y)}|,$$ where we have $\tilde{\gamma}:\s^2 \times \s^2 \rightarrow\s^2\times\s^2: (x,y) \mapsto (\gamma(x),\gamma(y))$, $J_\gamma(x)$ denoting the Jacobian matrix of $\gamma$ at x, and $| J_\gamma(x)|$ denoting its determinate. While the Q-transformation remains the same, the corresponding Q-function for $(f*\gamma)$ is given as $(q*\gamma) = (q\circ\tilde{\gamma})\sqrt{|J_\gamma(x)|}\sqrt{|J_\gamma(y)}|$. We denote the set of all such Q-functions as $\mathcal{Q}_{\s^2}$. To align two ConCon PDFs, we still use the objective function in Eq. (\ref{eqn: warpenergy}), but are searching for the optimal $\gamma^*$ from the diffeomorphisms in $\Gamma_{\s^2}$. 
    
    \subsection{Gradient Derivation}
    
    To derive the gradient $\nabla H$, we first extend Lemma \ref{dphi} for $\Omega = \s^2$.  
    
    \begin{lemma}\label{dphi_sphere}
    The differential of $\Phi \, : \, \Gamma_{\s^2} \mapsto [q]$ evaluated at the identity warp, $\gamma_{id}$, is a linear transformation $d\Phi_{\gamma_{id}} \, : \, T_{\gamma_{id}}(\Gamma_{\s^2}) \mapsto T_{q}([q])$, and is defined by the equation
    
    \begin{equation}
    \label{eqn:s2dev}
        \begin{aligned}
            (d\Phi_{\gamma_{id}}({b}))({x},{y}) &= {\begin{bmatrix}\frac{\partial q(x,y)}{\partial x} & \frac{\partial q(x,y)}{\partial y}\end{bmatrix}}{\begin{bmatrix} b(x) \\ b(y) \end{bmatrix}}  \\
            &\quad\quad\quad + \frac{q({x},{y})}{2}\Big( \nabla\cdot {b}({x}) + \nabla\cdot{b}({y})\Big),
        \end{aligned}
    \end{equation}
    
    \noindent where $\nabla \cdot {b}$ is the divergence of the vector field ${b} \in T_{\gamma_{id}}(\Gamma_{\s^2}).$
    \end{lemma}
    
    The proof is presented in Supplementary S4. With Lemma \ref{dphi_sphere}, we now can replace the directional derivative $\nabla_{{b}_i} H$  for $b_i \in T_{\gamma_{id}}(\Gamma_{\s^2})$ in Theorem \ref{th:dcost} with: 
    $$\nabla_{{b}_i} H = -2 \inner{q_1({x},{y}) - q_2(x,y)}{d\Phi_{\gamma_{id}}(b_i)},$$
    
        \noindent where we have $x$ and $y$ on $\s^2$, ${d\Phi_{\gamma_{id}}(b_i)}$ is defined in Eq. (\ref{eqn:s2dev}), $b_i$ is a vector field in $T_{\gamma_{id}}(\Gamma_{\s^2})$.

    \subsection{Orthonormal Basis for \(T_{\gamma_{\text{id}}}(\Gamma_{\mathbb{S}^2})\)}
    \label{sec:tangentbasis}
    
    To compute the differential of \(H\) via the approximation
    $\nabla H \approx \sum_{i=1}^{M} (\nabla_{b_i} H) \, b_i,$
    we require a basis for the tangent space $T_{\gamma_{id}}(\Gamma_{\s^2}) = \{b :  \mathbb{S}^2 \rightarrow T(\mathbb{S}^2) \; | b \text{ is a smooth tangent vector field on } \mathbb{S}^2\}$.
    Following the approach in \cite{RN26,RN25}, we construct an orthonormal basis for this space using spherical harmonics.
    
    Let \(Y_l^m : \mathbb{S}^2 \to \mathbb{C}\) denote the complex spherical harmonic of degree \(l\) and order \(m\) (with \(m = 0, \dots, l\)). These functions, which play a role analogous to the Fourier basis on the circle, form a complete orthonormal system for \(L^2(\mathbb{S}^2)\). In particular, each \(Y_l^m\) is defined as
    \[
    Y_l^m(\theta,\phi) = \sqrt{\frac{2l + 1}{4\pi}\frac{(l-m)!}{(l+m)!}}\, P_l^m(\cos\theta) \, e^{im\phi},
    \]
    where \(P_l^m\) denotes the associated Legendre polynomials.
    
    By taking the real and imaginary parts of \(Y_l^m\) (noting that \(\operatorname{Im}(Y_l^0)=0\)) for all \(l\) and \(m\), we obtain a collection of real-valued functions \(\{\psi_i(\theta,\phi)\}\) that form an orthonormal basis for \(L^2(\mathbb{S}^2)\). For example, an ordering might be:
    \[
    \psi_0 = Y_0^0,\quad \psi_1 = Y_1^0,\quad \psi_2 = \operatorname{Re}(Y_1^1),\quad \psi_3 = \operatorname{Im}(Y_1^1), \dots
    \]
    with the index \(i\) running over \(0,\dots,(l+1)^2-1\) (excluding the nonexistent imaginary parts for \(m=0\)).
    
    To change the function basis \(\{\psi_i\}\) to a basis for the tangent space \(T_{\gamma_{\text{id}}}(\Gamma_{\mathbb{S}^2})\), we use the gradient operator. By definition, the gradient of a scalar function on \(\mathbb{S}^2\) yields a tangent vector field. In standard spherical coordinates, with \(\theta \in [0,\pi]\) (the elevation) and \(\phi \in [0, 2\pi)\) (the azimuth), the gradient is expressed as
    $\nabla \psi_i(\theta, \phi) = \left[\frac{\partial \psi_i}{\partial \theta},\, \frac{1}{\sin \theta}\frac{\partial \psi_i}{\partial \phi}\right].$
     Next, we normalize these gradients:$
    \tilde{\eta}_i = \frac{\nabla \psi_i}{\|\nabla \psi_i\|}.$
    The set $\mathcal{B} = \{\tilde{\eta}_i\}$
    thus forms an orthonormal basis for the subspace of \(T_{\gamma_{\text{id}}}(\Gamma_{\mathbb{S}^2})\) consisting of gradient vector fields. However, this subspace accounts for only half of \(T_{\gamma_{\text{id}}}(\Gamma_{\mathbb{S}^2})\).
    
    To obtain the remaining half of the basis, we define a rotation operation. For any tangent vector \(v\) on \(\mathbb{S}^2\), let \(\ast v\) denote the vector obtained by rotating \(v\) counterclockwise by \(\pi/2\) (as seen from outside the sphere). Applying this to the gradients gives
    $
    \ast \nabla \psi_i(\theta,\phi) = \left[\frac{1}{\sin \theta}\frac{\partial \psi_i}{\partial \phi},\, -\frac{\partial \psi_i}{\partial \theta}\right].$
    After normalizing, we define
    $\tilde{\eta}_i^* = \frac{\ast \nabla \psi_i}{\|\ast \nabla \psi_i\|}.$
    The collection $\mathcal{B}^* = \{\tilde{\eta}_i^*\}$
    provides the remaining orthonormal basis elements. Hence, the union
    $\mathcal{B} \cup \mathcal{B}^*$
    forms a complete orthonormal basis for \(T_{\gamma_{\text{id}}}(\Gamma_{\mathbb{S}^2})\) \cite{RN26,RN25}.
    
      Finally, for any $b \in T_{\gamma_{id}}(\Gamma_{\s^2})$, the quantity $\nabla \cdot b$ is simply given by either $\nabla \cdot \frac{\nabla\psi_i}{\|\psi_i\|} = \frac{\nabla^2\psi_i}{\|\psi_i\|} = \frac{-l_i(l_i + 1)\psi_i}{{\|\psi_i\|}}$ for $ \tilde{\eta}_i \in \mathcal{B}$, where $-l_i(l_i + 1) \psi_i$ is from the Laplacian of the spherical harmonic $\psi_i$, or $0$ for $\tilde{\eta}^*_i \in \mathcal{B}^*$.

    \subsection{Computation and Implementation Details}
    \label{sec:comp_impl}
    
     Algorithm (\ref{alg:increamentalgamma}) now can be used to find the best warping to align two ConCon functions on $\s^2 \times \s^2$. However, there are still a few computational challenges to handle, e.g., how to represent the ConCon function $f$ in the code, how to apply the warping group action to $f$ to get $(f*\gamma)$, and how to compute the Jacobian of a given $\gamma$. In this subsection, we will introduce our solutions to these computational challenges. 
     
    {\bf Selecting Spherical Grid for Computation}. In practice, the ConCon function \( f \) and its \( Q \)-transformation \( q \) are realized only at a set of discrete locations \( \{ (x_i, x_j) \} \) for \( (x_i, x_j) \in \Omega \times \Omega \), with \( i, j = 1, 2, \ldots, V \). For \( \Omega = \mathbb{S}^2 \), it is necessary to define a grid over the sphere. We choose to generate this grid by recursively subdividing an icosahedron and projecting the vertices onto the unit sphere. This method of surface triangulation, known as an icosphere \cite{schneider2002geometric}, has several advantageous properties compared with the one used in \cite{RN25,RN26}:
    \begin{enumerate}
        \item Vertices on an icosphere are approximately evenly distributed. In other words, the density of vertices is no greater around the poles than any other part of the grid. This is different from uniform sampling on a polar coordinate system \cite{RN25,RN26}, and makes both deformation and interpolation easier.
        \item Vertices in lower resolutions icospheres are subsets of those in higher resolutions. As such, changing resolutions is easy, and it is possible to work with different resolutions from coarse to fine to get more accurate registrations.
    \end{enumerate}
    
    \noindent The number of subdivisions $G$ of the icosphere determines the number of vertices $V$ in the grid according to the formula $V = 4^G(10) + 2$. For example, when $G = 1$, the grid has $42$ vertices, when $G = 4$, there are $2562$ vertices, and when $G = 7$ there are $163,842$ vertices.  In this paper, we work with $G=4$ and refer to this grid as ico4. 
    
    {\bf Interpolation of ConCon Function $f$}.
    Given $f$ evaluated at $V^2$ discrete vertices on a grid, it will be necessary to evaluate $f(x,y)$ at any other point $(x,y) \in \s^2 \times \s^2$ using interpolation. For this, we use barycentric interpolation, an efficient algorithm that extends the concept of linear interpolation to higher-dimensional simplices. This interpolation is one of the most time-consuming steps in our algorithm. We have carefully designed an accurate and fast bi-variate interpolation presented in Supplementary S5.

     \textbf{Computation of \(\gamma\)'s Jacobian.}
    The Jacobian of the warping function \(\gamma\) is needed for evaluating the warping group action. Let $\gamma: \mathbb{S}^2 \to \mathbb{S}^2,$
    where \(\mathbb{S}^2\) is normally parameterized by spherical coordinates \((\theta,\phi) \in [0,\pi]\times [0,2\pi)\). However, we can use Cartesian coordinates or a vector field to represent $\gamma$. Actually, the Cartesian coordinate representation has many advantages, including that it avoids singularities when \(\sin\theta = 0\).
    We use function $\Psi$ to map between spherical and Cartesian coordinates:
    \begin{equation*}
      \label{eq:coord-transformations}
    \Psi(\theta,\phi)=
    \begin{pmatrix}
    \sin\theta\cos\phi\\
    \sin\theta\sin\phi\\
    \cos\theta
    \end{pmatrix},
    \Psi^{-1}(x,y,z)=
    \begin{pmatrix}
    \cos^{-1}(z)\\
    \tan^{-1}(y/x)
    \end{pmatrix}.
    \end{equation*}
    For a given $\gamma$ and a point $(\theta,\phi) \in \s^2$,  the warped point's spherical coordinate is given by $
    (\tilde{\theta},\tilde{\phi}) = \gamma(\theta,\phi).$ The Jacobian of \(\gamma\) is computed using a two-sided finite difference scheme in the tangent space \(T_{\Psi(\theta,\phi)}(\mathbb{S}^2)\) with its determinant given by:
    \[
    |J_{\gamma(\theta,\phi)}| = \sin\tilde{\theta}\,
    \left|\begin{matrix}
    \frac{\partial\tilde{\theta}}{\partial\theta} & \frac{1}{\sin\theta}\frac{\partial\tilde{\theta}}{\partial\phi} \\[1ex]
    \frac{\partial\tilde{\phi}}{\partial\theta} & \frac{1}{\sin\theta}\frac{\partial\tilde{\phi}}{\partial\phi}
    \end{matrix}\right|.
    \]
    {Details of the computation are provided in Supplementary S6.}

    \noindent \textbf{Spatial Derivative of \(q: \mathbb{S}^2 \times \mathbb{S}^2 \rightarrow \mathbb{R}_{\ge 0}\)}. One important quantity required in Lemma~\ref{dphi_sphere} is the directional derivative of \(q(x,y)\): ${\begin{bmatrix}\frac{\partial q(x,y)}{\partial x} & \frac{\partial q(x,y)}{\partial y}\end{bmatrix}}{\begin{bmatrix} b(x) \\ b(y) \end{bmatrix}}$.  This is computed using finite differences along predefined orthonormal bases \(\{w_x^{(1)}, w_x^{(2)}\}\) for \(T_x(\mathbb{S}^2)\) and \(\{w_y^{(1)}, w_y^{(2)}\}\) for \(T_y(\mathbb{S}^2)\). Representing \(b(x) = a_x^{(1)} w_x^{(1)} + a_x^{(2)} w_x^{(2)}\) and \(b(y) = a_y^{(1)} w_y^{(1)} + a_y^{(2)} w_y^{(2)}\), the total directional derivative is given by $
    dq_{x}\bigl(w_x^{(1)},y\bigr)a_x^{(1)} + dq_{x}\bigl(w_x^{(2)},y\bigr)a_x^{(2)} + dq_{y}\bigl(x,w_y^{(1)}\bigr)a_y^{(1)} + dq_{y}\bigl(x,w_y^{(2)}\bigr)a_y^{(2)},$
    where \(dq_x\) and \(dq_y\) are approximated via the exponential map and barycentric interpolation with a small \(\epsilon\). Details are provided in Supplementary S7.

    \section{Extension to $\Omega = \s_1^2 \cup \s_2^2$}
    \label{sec:extens2us2}
    In our application, the domain of interest for cortical surface registration contains two hemispheres. As such, the tools developed above need to be expanded to work with $\Omega = \Omega_1 \cup \Omega_2$, where $\Omega_1 \cap \Omega_2 = \emptyset$, and $\Omega_1$ denotes the left hemisphere and $\Omega_2$ denotes the right hemisphere. The diffeomorphism $\gamma$ can now be written as:
    \begin{equation*}
        \gamma(x) = \begin{cases}
            \gamma^1(x) & \text{if } x \in \Omega_1 \\
            \gamma^2(x) & \text{if } x \in \Omega_2.
    \end{cases}
    \end{equation*}
    
    \noindent And we can redefine the mapping $\Phi$ in Eq. (\ref{def:phi}) as: $ \Phi(\gamma)(x,y) =  $
    \begin{equation}\label{eq:full_phi}
    \begin{cases}
            q(\gamma^1(x),\gamma^1(y))\sqrt{|J_{\gamma^1}(x)|}\sqrt{|J_{\gamma^1}(y)|}, & \text{if } x \in \Omega_1, y \in \Omega_1 \\
            q(\gamma^2(x),\gamma^1(y))\sqrt{|J_{\gamma^2}(x)|}\sqrt{|J_{\gamma^1}(y)|}, & \text{if }x \in \Omega_2, y \in \Omega_1 \\
            q(\gamma^1(x),\gamma^2(y))\sqrt{|J_{\gamma^1}(x)|}\sqrt{|J_{\gamma^2}(y)|}, & \text{if }x \in \Omega_1, y \in \Omega_2 \\
            q(\gamma^2(x),\gamma^2(y))\sqrt{|J_{\gamma^2}(x)|}\sqrt{|J_{\gamma^2}(y)|}, & \text{if }x \in \Omega_2, y \in \Omega_2.
    \end{cases}
    \end{equation}
    
    \noindent Assuming that $q(x,y)$ is continuous and differentiable when $x$ and $y$ are in different sub-domains, then each of the cases in Eq. \eqref{eq:full_phi} can be solved as before. As such, the theory remains unchanged. In finding the optimal warping function to align ConCon functions, we propose an iterative procedure where the warp for each hemisphere is updated individually, i.e., we first fix $\gamma^2$ and derive the gradient w.r.t. $\gamma^1$.  The directional derivative of $H$ in Eq. (\ref{dcost}) w.r.t. $\gamma^1$ in the direction ${b}_i \in T_{\gamma_{id}}(\Gamma_{\s^2})$ is carefully derived in Supplementary S8. Similarly, we can derive the directional derivative of $H$ w.r.t. $\gamma^2$ with $\gamma^1$ fixed. Algorithm \ref{alg:gamma} presents a detailed procedure to align $f_2$ to $f_1$ on $\Omega \times \Omega$ with their Q-transformations $q_1$ and $q_2$, respectively. 
    
    \begin{algorithm}
    \caption{Align ConCons for $\Omega = \Omega_1 \cup \Omega_2$}\label{alg:gamma}
    \begin{algorithmic}
    \State \textbf{Input}: $f_1, f_2$ and their Q-transformations $q_1$ and $q_2$
    \State Set $k = 0$, $\gamma^{1(0)} = \gamma^{2(0)} = \gamma_{id}$, a small constant $\epsilon$ and step size $\delta$
    \vspace{0.2em}
    \While{$||\nabla H_{\gamma^1}|| > \epsilon$ and $||\nabla H_{\gamma^2}|| > \epsilon$}
    \State ${q}^\prime_2 \gets (q_2 * \gamma^{(k)})$ where $\gamma^{(k)}=(\gamma^{1(k)},\gamma^{2(k)})$
    \vspace{0.2em}
    \State \text{Compute} $\nabla H_{\gamma^1} = \sum_{i=1}^M(\nabla_{b_i}H_{\gamma^1})b_i$ using Eq. S.8 based on $q_1$ and $q_2^\prime$. And compute $\nabla H_{\gamma^2}$ similarly.
    \vspace{0.2em}
    \State \text{Compute incremental warp}  ${\gamma}^{1}_k \gets \exp_{\gamma_{id}}({\delta \nabla H_{\gamma^1}})$ and similarly ${\gamma}^{2}_k$
    \State \text{Update overall warp}  $\gamma^{1(k + 1)} \gets \gamma^{1(k)} \circ {\gamma}^{1}_k$ and similarly $\gamma^{2(k + 1)}$.
    \vspace{0.2em}
    \State $k \gets k + 1$
    \EndWhile
    \end{algorithmic}
    \vspace{0.2em}
    \noindent \textit{Here $\exp_{\gamma_{id}}$ is the exponential map on $\Gamma_{\s^2}$ evaluated at $\gamma_{id}$.}
    \end{algorithm}
    
    \section{Defining a Template for Population-based Registration}
    \label{sec:deftemp}
    The final ingredient for a full registration algorithm is the definition of a template. The goal of this registration framework is to align the ConCon functions $f_1,...,f_N$ in such a way that similar features are matched as closely as possible between functions. In order to do this, we need to find a mean function, or `template', that minimises the warping necessary to bring each of the functions into alignment. We will approach this via two steps: (1) given a set of $N$ connectivity functions $\{f_j \in \mathcal{F} \; | \; j = 1,2,\dots,N \}$ and their $Q$-mapped representations $\{q_j \in \mathcal{Q} \; | \; j = 1,2,\dots,N \}$, we find the mean $[\mu]$ of the orbits $\{[q_j]\}$; (2) We find the center element in $[\mu]$ w.r.t. $\{q_j\}$ and select it as the template. However, if we suspect that the sample of connectivity functions includes some outliers, we can choose to use the median of $\{[q_j]\}$ instead of the mean \cite{RN79,RN80}. In Supplementary S9, we present the full ENCORE algorithm with the following three modules: 1) finding $[\mu]$, 2) finding the center of $[\mu]$ w.r.t. $\{q_j\}$ and 3) a complete algorithm for ConCon alignment.

    \section{Simulation Study}
    \label{sec:simulation}
    
    \begin{figure}[t!]
        \centering
    \includegraphics[width=\columnwidth]{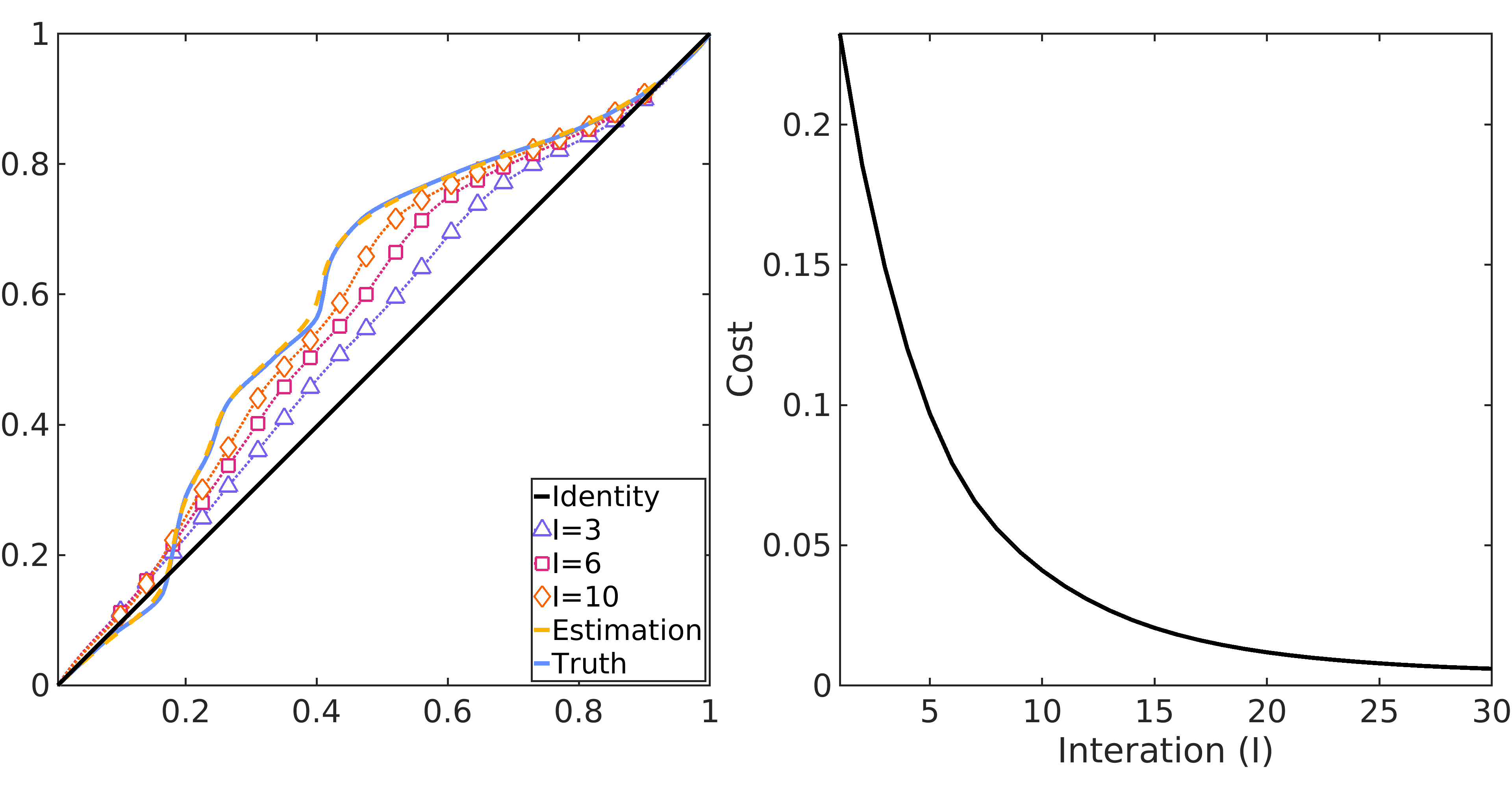}
        \caption{\small{The left panel shows the resulting $\hat{\gamma}$ after iteration I=3, I=6, I=10, after converging, and the true $\gamma$ for a random simulated case for $\Omega = [0,1]$. For context, the $\LL^2$ norm between $\gamma$ and $\hat{\gamma}$ for each iteration is $0.086$, $0.054$, $0.033$, $0.005$, respectively, and $0.136$ between $\gamma$ and identity. The right panel shows the corresponding cost function after each iteration.}}
        \label{fig:simulation_fig}
    \end{figure}
    
    To demonstrate the efficacy of Algorithm \ref{alg:increamentalgamma}, we conduct a simulation study for $\Omega = [0,1]$. The one-dimensional case is simpler to visualize and provides helpful intuition regarding how registration works in more complicated domains.
    
    {\bf Simulate Warping Functions}. We simulate $\gamma \in \Gamma_{[0,1]}$ in the following way: 1) simulate  $\dot{\gamma}(x_k) \sim \text{Uniform}(0,1)$ at uniformly spaced points $x_k \in [0,1]$ for $k = 1,2,\dots,K$, 2) take the normalized cumulative sum of the discrete observations $\dot{\gamma}(x_k)$ to generate a discrete $\gamma$, i.e., $\gamma(x_k) = \sum_{i=1}^k \dot{\gamma}(x_i)$, and 3) finally, we make $\gamma(x)$ continuous through shape-preserving piece-wise cubic interpolation \cite{ma2022stochastic}. Panel B of Fig. \ref{fig:warp} shows an example of a simulated warping function. 
    
    {\bf Simulate ConCon Functions}. To simulate ConCon functions, we start with simulating points on $[0,1] \times [0,1]$ and then apply the kernel density estimation method to estimate the underlying ConCon functions $f$. More specifically, let $P_i = \left\{\left(\frac{i-1}{p},\frac{i}{p}\right]\right\}$ be the elements of the discrete parcellation of the domain $\Omega = [0,1]$ into $p = 200$ distinct parcels. Then, analogous to real brain structural connectivity, we generate a set of point pairs $\{(u_j,v_j)\}$, for $j = 1,2,\dots,20,000$, with $(u_j,v_j)$ representing the two ending points of the $j$-th fiber connection. Panel A of Fig. \ref{fig:warp} shows one example of a simulated ConCon function and Panel C shows an example of the warped Concon. We simulate N=$10,20,40,80$ such ConCon functions using the parameters given in Supplementary Table S1, apply the simulated $\gamma$s, then run our entire ENCORE pipeline to estimate the $\hat{\gamma}^{-1}$ functions that recover the originally simulated ConCon functions. We then evaluate the results using the $\LL^2$ norm $(\int_0^1 |\gamma(t) - \hat{\gamma}(t)|^2 dt)^{1/2}$. Fig. \ref{fig:simulation_fig} shows the evolution of $\hat{\gamma}$ when running ENCORE on one of the randomly generated ConCon functions. Further examples of warping functions and their estimates are shown in Supplementary Fig. S1. The results of the simulation study, as shown in Table \ref{tbl:simulation}, demonstrate that our method is able to reliably estimate the warping functions required for accurate registration.
    
    \begin{table}[!ht]
    \caption{Simulation Results}
    \centering
    \begin{tabular}{|l|c|c|c|c|}
    \hline
     & \textbf{N=10} & \textbf{N=20} & \textbf{N=40} & \textbf{N=80} \\ \hline
    \textbf{A} & \textbf{0.059} (0.005) & \textbf{0.058} (0.007) & \textbf{0.057} (0.008) & \textbf{0.058} (0.014) \\ \hline
    \textbf{B} & 0.251 (0.062) & 0.226 (0.043) & 0.214 (0.051) & 0.221 (0.051) \\ \hline
    \end{tabular}
    \label{tbl:simulation}
    \caption*{\small{Population mean (standard deviation in parenthesis) of the $\LL^2$ norm between, (\textbf{A}) $\gamma$ and $\hat{\gamma}$, and (\textbf{B}) $\gamma$ and $\gamma_{id}$, for varying population size.}}
    \end{table}
    
    \section{Real Brain Connectome Data Alignment}
    \label{sec:realdata}
    \subsection{Dataset and Competing Methods}
    \label{sec:datacompmethod}
    We mainly used data from the Human Connectome Project (HCP), which is a milestone dataset for mapping human brain connectomics. The data used in this paper were obtained from the ConnectomeDB website, which consists of preprocessed T1-weighted and diffusion MRI (dMRI) images from the HCP Young Adult dataset. Information about image acquisition and dMRI preprocessing steps can be found in \cite{glasser2013minimal}. Briefly, imaging was conducted on the 3T Siemens Connectome scanner in Erlangen, Germany. High-resolution T1-weighted anatomical images were obtained at 0.7 mm isotropic resolution. Diffusion imaging was performed with 90 diffusion directions at three nonzero b-values (1,000, 2,000, and 3,000 s/mm$^2$) and 1.25 mm isotropic resolution.
    
    The data were then processed using the SBCI pipeline \cite{cole2021}, a state-of-the-art dMRI preprocessing pipeline. SBCI constructs cortical surfaces and their spherical parameterizations via FreeSurfer \cite{RN21}. It then estimates white matter fiber tracts, commonly referred to as streamlines in the literature, connecting cortical surfaces using surface-enhanced tractography \cite{RN15}. On average, approximately 500,000 streamlines were generated per subject. Fig. \ref{fig:Figure1}, panel B, shows one example of the streamline endpoints $O_i$. Finally, a point process model in \cite{cole2021} was used to estimate the intensity function $f^*$ (defined in Eqn. (\ref{eqn:first_moment_assumption})) from $O_i$, alongside the normalized ConCon function $f$. These ConCon functions are represented as $5124 \times 5124$ adjacency matrices based on the ico4 spherical grid described in Section \ref{sec:comp_impl}.

    We used the ENCORE algorithm presented in Section \ref{sec:deftemp} to first compute a template ConCon from 200 randomly chosen subjects. The computed template is displayed in Supplementary Fig. S2 panel (A), which is used as the template to align other ConCon functions. For each ConCon $f$, we obtained warped ConCon functions ($f*\gamma$) along with the warping functions $\gamma$ for the left and right hemispheres. Both ($f*\gamma$) and $\gamma$ are stored for further analysis.
    
     In this study, we compare ENCORE with several established cortical surface alignment techniques: {\bf MSMAll} \cite{RN5}, {\bf MSMSulc} \cite{RN5}, {\bf FreeSurfer} \cite{RN21}, and {\bf S3Reg} \cite{RN30}. MSMAll (Multimodal Surface Matching All)  aligns cortical surfaces using a combination of multimodal features, such as sulcal depth, curvature, and myelin maps. MSMSulc, a variant of MSM, relies solely on sulcal depth to drive alignment. FreeSurfer performs surface registration using features like cortical thickness and curvature and is widely used for its robust preprocessing and anatomical accuracy. S3Reg is a deep learning method, and it aligns cortical surfaces across individuals using univariate features at each vertex. Since S3Reg (Superfast Spherical Surface Registration) operates on univariate features, we performed registrations separately using sulcal depth and cortical thickness as inputs. Additionally, we derived a novel feature, TotalConn, defined as the sum of connectivity values at each vertex (i.e., the row sum of the ConCon matrix), and used it for registration. 
    
    \subsection{Computational Cost and Robustness Analysis}
    We implemented the ENCORE algorithm using a combination of MATLAB and C++, together with the publicly available libigl C++ libraries. With careful use of vectorization to eliminate all but essential loops from the code, strict memory management (accessing array elements directly through pointers and in contiguous blocks) within the C++ routines, and libigl data structures, the code runs at a competitive speed to other registration methods. On average, using an Intel CPU (2.50 GHz) with 5GB of RAM, it takes 40 seconds to register a pair of ConCon with the ico4 grid. It takes approximately 20 minutes to calculate a template from 200 ConCon functions. The template estimation procedure could be made quicker, but it is currently slow due to the necessity to load and unload connectomes individually during the iterative process in order to keep the memory demands reasonably low.

\begin{figure}[t!]
        \centering
        \includegraphics[width=\columnwidth]{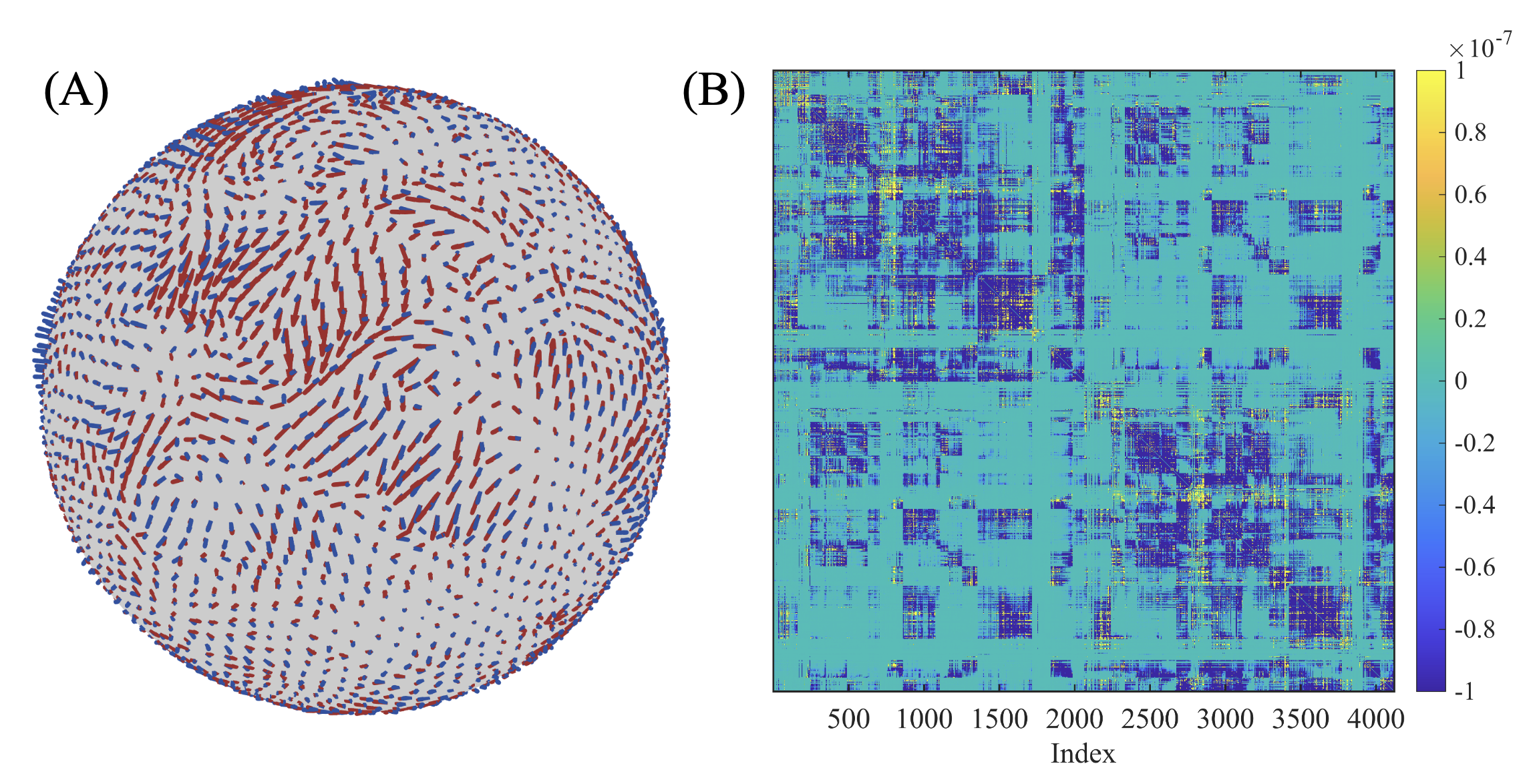}
        \caption{\small{Panel (A): Two warping functions (displayed here as vector fields) for the left hemisphere generated by the ENCORE method using two templates (see Figure S2) for a randomly selected HCP subject. Each template is built based on 200 randomly selected subjects. Panel (B): Element-wise difference in variance between ENCORE and MSMAll (ENCORE - MSMAll) for 200 randomly selected HCP subjects.}}
        \label{fig:vector_field_comparison}
        \label{fig:vardiff}
    \end{figure}

    We also tested the robustness of the template-building and alignment process. We ran the template-building process with another 200 randomly selected subjects from HCP and obtained a different template. The second template is displayed in Supplementary Fig. S2 panel (B), and one can see that it looks very similar to the first template. Moreover, we randomly selected one HCP  subject and aligned the subject to both templates. Fig. \ref{fig:vector_field_comparison} panel (A) shows the $\gamma^{1}$'s (warping function for the left hemisphere) obtained by using the two templates. Here, the warping functions are displayed as vector fields, and we can see that the two vector fields are very similar to each other, indicating the robustness of the template-building and ConCon alignment processes in ENCORE.  We also studied the angle between the vector fields, considering only vectors with amplitudes above the average. Across the left hemispheres of 100 randomly selected subjects, the mean angle is 20.22 degrees with a standard deviation of 2.85 degrees.

    \subsection{Exploration of Aligned Structural Connectomes} 
    \label{sec:eda_concon}
    
    When ConCon functions are well aligned, we expect their variance to be reduced. Here, we compared the cross-sectional (element-wise) variance for MSMAll and ENCORE methods with $200$ randomly selected subjects. Note that according to the validation results in Section \ref{sec:valid}, MSMAll is the best method among all competitors in aligning major fiber bundle's endpoints. Limited by space, we present results comparing ENCORE to MSMAll in this and the next subsections.  Fig. \ref{fig:vector_field_comparison} panel (B) shows the variance difference between ENCORE and MSMAll  (ENCORE - MSMAll). By examining the result, we see that in most regions, the connectome variance gets reduced compared with the non-connectivity-based alignment method (i.e., MSMAll), demonstrating that ENCORE is more effective in aligning the ConCons. 
    
\begin{figure}[t!]
        \centering
        \includegraphics[width=0.8\columnwidth]{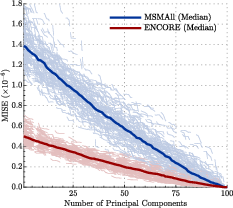}
        \caption{{\small Reconstruction error of structural connectomes registered with MSMAll and ENCORE in 100 HCP participants. Dashed lines represent individual reconstruction errors, while solid lines denote the median error.}}
        \label{fig:reconstruction_error}
    \end{figure}

    
    In another experiment, we compared ENCORE with MSMAll regarding the efficiency of connectome representation after registration. Here, we employed a conventional Functional Principal Component Analysis (FPCA) method. For 100 randomly selected ConCons, FPCA was performed separately on ConCons registered with MSMAll and those aligned using ENCORE. We then reconstruct the connectomes using an increasing number of PCs and quantify the reconstruction error—defined as the mean integrated squared error between the original and reconstructed connectomes for each subject. Fig. \ref{fig:reconstruction_error} presents the reconstruction error versus the number of PCs. A lower reconstruction error using fewer PCs indicates that the leading components capture a greater proportion of meaningful information. Our results demonstrate that connectomes registered with ENCORE exhibit significantly lower reconstruction errors than those registered with MSMAll. This enhanced efficiency implies that fewer PCs are needed to explain more variance, thereby benefiting downstream analyses and potentially reducing unwanted noise in the data.
    
    \subsection{Exploration of Estimated  Diffeomorphisms}
    \label{sec:eda_diffeo}
    
\begin{figure}[t!]
        \centering
    \includegraphics[width=\columnwidth]{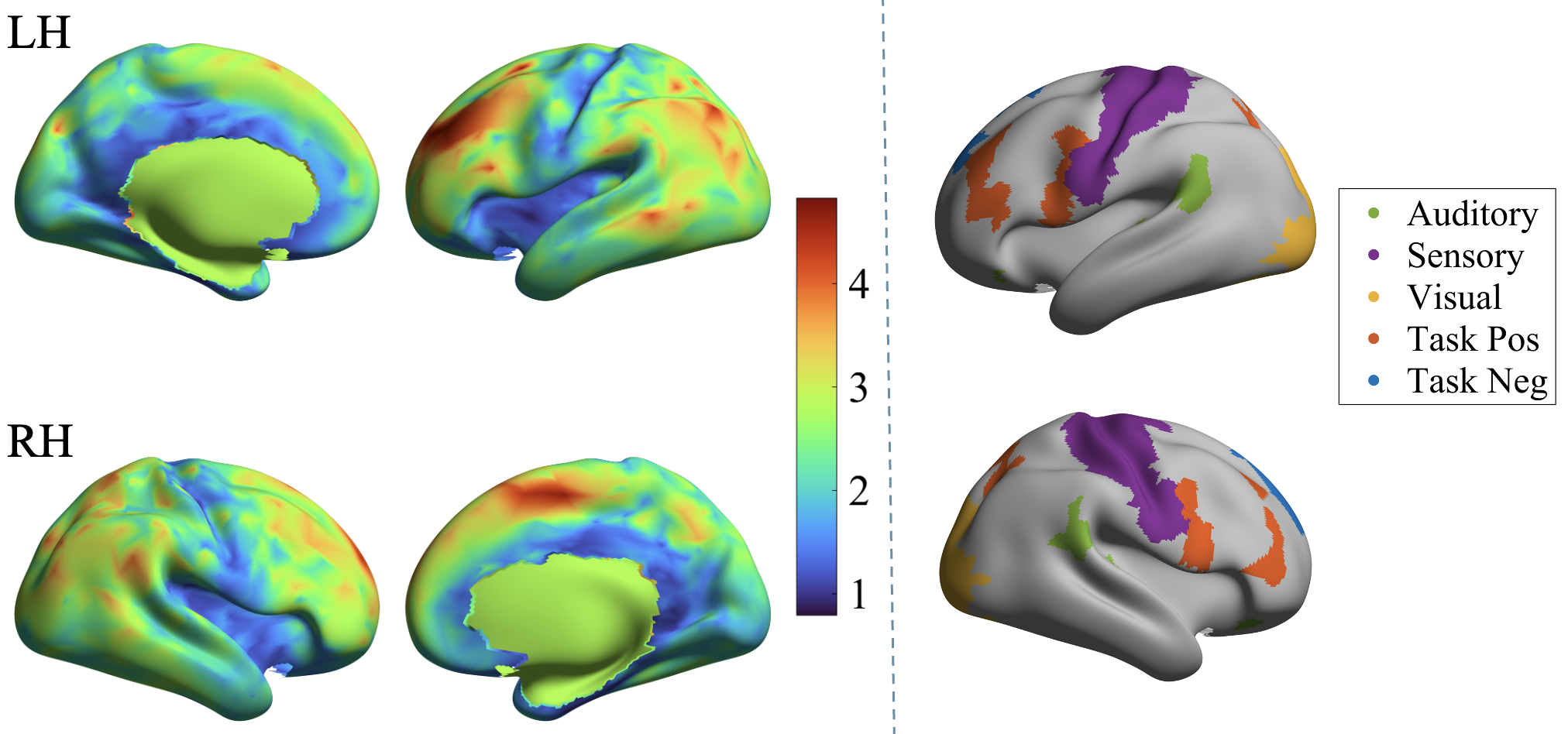}
        \caption{\small{The Left panel shows the mean magnitude of deformation (measured as the mean distance each vertex is translated by the warping functions), and the right shows the regions of interest.}}
        \label{fig:mean_deformation}
    \end{figure}

    The warping functions produced by ENCORE provide insight into the amount of energy required to align brain networks across subjects. A larger deformation indicates poor alignment of specific regions, while a smaller deformation suggests that connectivity is already well aligned. In this study, we aim to investigate the areas where ENCORE expends more or less energy during alignment. Given that warping functions encode alignment deformation information, we present the average magnitude of deformation across 200 randomly selected subjects. Fig. \ref{fig:mean_deformation} shows the mean translation distance for each vertex resulting from the warping functions applied to the population to align to the template. 
    
    The results reveal distinct patterns of deformation across the cortex, providing insight into the inter-individual variability in brain network alignment. Our results indicate that task-related regions, including task-positive and task-negative networks, exhibit significantly larger deformations compared to sensory and motor areas. This suggests that higher-order cognitive regions, which are involved in executive function, attention, and internally directed thought, require more extensive transformations to achieve alignment across individuals. The high deformation in these regions likely reflects their greater anatomical and functional variability, shaped by developmental trajectories and experience-driven plasticity \cite{frost2012measuring,mueller2013individual}. In contrast, sensory cortices, such as the auditory and visual regions, display smaller deformation magnitudes, reinforcing the idea that these areas are evolutionarily conserved and exhibit stable functional architectures across individuals due to genetic and environmental constraints \cite{RN18,gordon2017precision}. Additionally, hemispheric differences in deformation magnitudes suggest that functional specialization may influence alignment variability, with the left hemisphere showing slightly greater deformations in frontal lobe regions. The observed deformation patterns provide new insights into the neurobiological constraints shaping cortical connectivity and have broader implications for understanding individual differences in cognition, neurodevelopment, and neurological disorders.
    
    \begin{figure}[t!]
         \centering
        \includegraphics[width=\columnwidth]{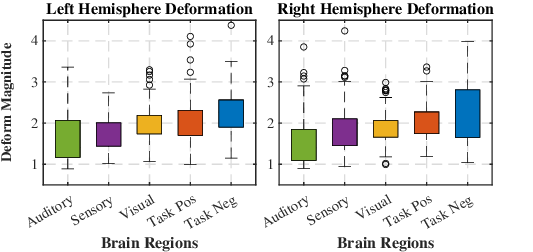}
        \caption{{\small Shows the magnitude of deformation in each parcellation area (measured as the distance each vertex in the parcellation area is translated by the warping functions)}}
        \label{fig:parcellation}
    \end{figure}
    
    \subsection{Validation of ENCORE Alignment}
    \label{sec:valid}
    To validate the accuracy of ENCORE's alignment, we assess whether the endpoints of major fiber bundles become better aligned after applying the method. Specifically, we evaluate the overlap of fiber bundle endpoints across subjects and compare the performance of ENCORE to other commonly used registration methods. Our hypothesis is that well-registered connectomes will exhibit higher endpoint agreement across subjects, as measured by the Dice similarity coefficient.
    
    To achieve this, we use RecoBundlesX, a multi-atlas white matter bundle segmentation framework, to extract major fiber bundles from individual subjects. RecoBundlesX is an enhanced version of the Recobundles algorithm \cite{garyfallidis2018recognition}, incorporating multi-parameter fusion to improve bundle segmentation accuracy and robustness. This approach allows for more precise and reliable extraction of white matter pathways across subjects. The segmentation was performed using the Scilpy \cite{st2023bundleseg} pipeline.
    
    For validation, we selected 50 randomly chosen subjects and extracted five major fiber bundles: bilateral Middle Longitudinal Fasciculus (MdLF), Inferior and Superior Longitudinal Fasciculus (ILF, SLF), Frontal Aslant Tract (FAT), and Anterior Cingulate Cortex (CG\_An). Supplementary Fig. S3 shows these fiber bundles from one HCP subject. These bundles were chosen based on their representation of key brain pathways and their sufficient streamline density, which allows for a reliable computation of Dice coefficients. We calculated the Dice coefficient to quantify endpoint overlap for each fiber bundle across all subject pairs after applying different alignment techniques. This resulted in seven 50 × 50 upper triangular matrices of Dice scores, corresponding to different registration methods, including MSM-based methods (e.g., MSMsulc using sulcal depth and MSMall using multiple cortical features), S3Reg-based methods (e.g., S3RegSulc using sulcal depth, S3RegThickness using cortical thickness, and S3RegTotalConn using total connectivity), FreeSurfer registration, and ENCORE. Among these methods, S3RegTotalConn utilizes the total connectivity derived from ConCon at each vertex as the registration feature.
    
\begin{figure}[t!]
        \centering
        \includegraphics[width=0.8 \columnwidth]{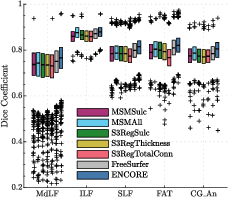}
        \caption{\small{Shows the Dice coefficients for five major fiber bundles across different registration methods.}}
        \label{fig:dice_score}
    \end{figure}

    Fig. \ref{fig:dice_score} presents the Dice scores for the selected fiber bundles across different registration methods. The results indicate that ENCORE consistently outperforms other methods, yielding higher Dice scores across all bundles. Notably, CG\_An, SLF, and MdLF exhibit the largest improvements, highlighting ENCORE’s ability to align higher-order association pathways, which are known for their high inter-individual variability due to cognitive specialization and developmental plasticity. The CG\_An, a key hub in the default mode network, the SLF, critical for attention and language, and the MdLF, involved in auditory and language processing, all show improved alignment, reducing anatomical discrepancies. By better preserving these complex connections, ENCORE enhances group-level comparisons in neuroimaging studies and improves the reliability of structural connectivity analyses in both health and disease.
    
    \subsection{Statistical Analysis of Aligned ConCon Functions}
    To evaluate the effectiveness of ENCORE for brain connectivity's downstream statistical analysis, we assess its ability to improve trait prediction using ConCon data from the HCP. Specifically, we focus on two cognitive traits: Oral Reading Recognition and the Picture Vocabulary Test, both of which assess language-related abilities and are influenced by brain connectivity patterns. The goal is to compare ENCORE’s predictive performance against other commonly used registration methods for connectivity data.  
    
    For each trait, we randomly selected 400 subjects and obtained aligned connectivity matrices using different registration methods. Additionally, we also exprimented to incorporate information from the estimated warping functions in ENCORE to examine its contribution to prediction accuracy. We employ Principal Component Regression (PCR) as our predictive framework for simplicity, where the model's tuning parameters were determined using  5-fold cross-validation in the 70\% training data. The selected model was evaluated based on the 30\% testing data. Performance was evaluated based on the correlation between observed and predicted values in 100 repeated runs. 
    
\begin{figure}[htbp]
        \centering

            \includegraphics[width=0.8\columnwidth]{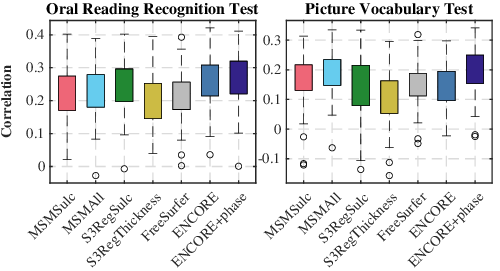}
            \vspace{2mm} 
        \parbox{\columnwidth}{
            \centering
            (a) \hspace{0.45\columnwidth} (b)
        }
        \caption{\small{Prediction results for the Oral Reading Recognition Test (a) and the Picture Vocabulary Test (b) using different registration methods.}}
        \label{fig:prediction}
    \end{figure}

    Fig. \ref{fig:prediction} illustrates the prediction results. We can see that ENCORE can enhance predictive accuracy compared to other registration methods for these traits.  When the warping information was included, the prediction could be further improved, indicating that the warping function also carries information about these traits. These results suggest that ENCORE not only improves connectome alignment but also enhances the ability to extract meaningful individual differences in cognitive traits, reinforcing its value for downstream statistical analyses in neuroimaging research.
    
    \section{Conclusion and Discussion}
    \label{sec:conclusiondis}
    
    In this work, we introduced ENCORE (mEtric-based coNtinuous COnnectivity REgistration), a novel framework for aligning brain structural connectivity using a continuous connectivity representation. Unlike traditional atlas-based methods that rely on predefined ROIs and neglect connectivity patterns during alignment, ENCORE formulates brain network registration as an optimal diffeomorphism problem, directly aligning high-resolution connectivity profiles. By employing a square-root transformation and an isometry-preserving group action under the \(\mathbb{L}^2\) metric, we developed an iterative algorithm for optimizing the alignment. We validated the theoretical foundations of ENCORE by first testing it on a simplified 1D domain (\(\Omega = [0,1]\)) before extending it to manifold domains (\(\Omega = \mathbb{S}^2_1 \cup \mathbb{S}^2_2\)) for real-world applications.  
    
     Applied to HCP data, ENCORE consistently reduced inter-subject connectivity variability, improved the efficiency of connectome representation, and enhanced the alignment of major fiber bundle endpoints, as evidenced by higher Dice similarity coefficients. Additionally, the estimated warping functions provided insight into inter-individual variability, with greater deformations observed in task-related cognitive regions compared to sensory-motor areas, highlighting the neurobiological basis of connectivity differences. In downstream statistical analyses, ENCORE improved trait prediction accuracy for language-related cognitive measures, further demonstrating its potential for neuroimaging applications.  
    
    An efficient implementation of ENCORE is available at \url{https://github.com/sbci-brain/ConCon_Alignment}. While ENCORE shares conceptual similarities with MSM and FreeSurfer, it offers key advantages: (1) alignment based on high-resolution connectivity patterns, (2) inverse consistency, and (3) strict diffeomorphic warping functions—all of which are critical properties in image registration. Given its simple cost function, the ENCORE framework is well-suited for extension to a deep learning (DL)-based approach to accelerate inference. However, challenges remain in training DL models to handle high-resolution connectivity warping and ensuring that outputted warping functions remain diffeomorphic.

\bibliography{references}

\end{document}


\title{\bf Supplemental Materials}
\author{}
\date{}
\maketitle
\setcounter{page}{13}
\makeatletter
\renewcommand{\theequation}{S.\arabic{equation}}
\renewcommand{\thefigure}{S\arabic{figure}}
\renewcommand{\thesection}{S\arabic{section}}

\section{Proof of Lemma II.1}

\begin{proof}
    The proof is straightforward. On the left, we have $d_{\LL^2}(q_1,q_2)=\cos^{-1}(\iint q_1(x,y)q_2(x,y) dxdy)$. And on the right, 
    \begin{equation} \label{eq1}
\begin{split}
d_{\LL^2}((q_1*{\gamma}),(q_2*{\gamma})) &=  \cos^{-1}(\iint (q_1*{\gamma})(q_2*{\gamma})dxdy) \\
 & = \cos^{-1}( \iint (q_1 \circ \tilde{\gamma})(q_2 \circ \tilde{\gamma})\dot{\gamma}(x)\dot{\gamma}(y)dxdy) \\
 & = \cos^{-1}( \iint q_1(\gamma(x),\gamma(y))q_2(\gamma(x),\gamma(y))\dot{\gamma}(x)\dot{\gamma}(y)dxdy ) \\
 & = \cos^{-1}(\iint q_1(\tilde{x},\tilde{y})q_2(\tilde{x},\tilde{y})d\tilde{x}d\tilde{y}) \\
 & = d_{\LL^2}(q_1,q_2).
\end{split}
\end{equation}
\end{proof}

\section{Proof of Lemma II.2}

\begin{proof} Let $\bar{\gamma}_p \, : \, [-\epsilon,\epsilon] \times [0,1] \rightarrow [0,1]$, denoted as $\bar{\gamma}_p(t, x)$, be  a differentiable path in $\Gamma_{[0,1]}$, passing through $\gamma_{id}$ at time $t = 0$. Furthermore, let the velocity of $\bar{\gamma}_p$ at $t = 0$ be given by the tangent velocity function ${b} \in T_{\gamma_{id}}(\Gamma_{[0,1]})$. As such, the path satisfies the initial conditions:
\begin{equation*}
    \bar{\gamma}_p(0, {x}) = \gamma_{id}({x}),\; 
    \frac{\partial \bar{\gamma}_p(t, {x})}{\partial {x}}\bigg|_{t=0} = 1,\;
    \frac{\partial \bar{\gamma}_p(t, {x})}{\partial {t}}\bigg|_{t=0} = {b}({x}),\;
    \frac{\partial \bar{\gamma}_p(t, {x})}{\partial {x}\partial t}\bigg|_{t=0} = \dot{b}({x}).
\end{equation*}

\noindent Since $\bar{\gamma}_p(t,\cdot)$ is a path in $\Gamma_{[0,1]}$, $\Phi(\bar{\gamma}_p(t,\cdot))$ is the corresponding path in $[q]$. Furthermore, since the initial velocity of $\bar{\gamma}_p$ is ${b}$, the initial velocity of the corresponding path in $[q]$ is $d\Phi_{\gamma_{id}}({b})$. And so, let $u=\bar{\gamma}_p(t,x)$, $v=\bar{\gamma}_p(t,y)$, differentiating $\Phi$ with respect to $t$ gives
\begin{align*}
    \frac{d}{dt}\Phi(\bar{\gamma}_p)({x},{y}) &= \frac{d}{dt}\left(q\big(\bar{\gamma}_p(t,{x}),\bar{\gamma}_p(t,{y})\big) \sqrt{\frac{\partial \bar{\gamma}_p(t,{x})}{\partial {x}}}\sqrt{\frac{\partial \bar{\gamma}_p(t,{y})}{\partial {y}}}\right) \\ 
    &= \frac{\partial q(
    u,v)}{\partial u}\frac{\partial \bar{\gamma}_p(t,x)}{\partial t}\sqrt{\frac{\partial \bar{\gamma}_p(t,{x})}{\partial {x}}}\sqrt{\frac{\partial \bar{\gamma}_p(t,{y})}{\partial {y}}} \\
    &\hspace{1em} + \frac{\partial q(
    u,v)}{\partial v}\frac{\partial \bar{\gamma}_p(t,y)}{\partial t}\sqrt{\frac{\partial \bar{\gamma}_p(t,{x})}{\partial {x}}}\sqrt{\frac{\partial \bar{\gamma}_p(t,{y})}{\partial {y}}} \\
    &\hspace{1em}+ \frac{1}{2 \sqrt{\frac{\partial \bar{\gamma}_p(t,{x})}{\partial {x}}}}\left(q\big(\bar{\gamma}_p(t,{x}),\bar{\gamma}_p(t,{y})\big) \frac{\partial \bar{\gamma}_p(t,{x})}{\partial {x} \partial t}\sqrt{\frac{\partial \bar{\gamma}_p(t,{y})}{\partial {y}}}\right) \\
    &\hspace{1em}+ \frac{1}{2 \sqrt{\frac{\partial \bar{\gamma}_p(t,{y})}{\partial {y}}}}\left(q\big(\bar{\gamma}_p(t,{x}),\bar{\gamma}_p(t,{y})\big) \sqrt{\frac{\partial \bar{\gamma}_p(t,{x})}{\partial {x}}}\frac{\partial \bar{\gamma}_p(t,{y})}{\partial {y} \partial t}\right)
\end{align*}

\noindent via the chain rule. Substituting in all the initial conditions, gives
\begin{align*}
\frac{d}{dt}\Phi(\bar{\gamma}_p)(x,y)|_{t = 0} &= 
\frac{\partial q\left({x},{y}\right)}{\partial x} b(x) + \frac{\partial q\left({x},{y}\right)}{\partial y} b(y) \\
&\hspace{1em} + \frac{1}{2}(q\left(\gamma_{id}({x}),\gamma_{id}({y})\right) \dot{b}(x) + \frac{1}{2}(q\left(\gamma_{id}(x),\gamma_{id}({y})\right) \dot{b}(y) \\
    &= \begin{bmatrix}\frac{\partial q(x,y)}{\partial x} & \frac{\partial q(x,y)}{\partial y}\end{bmatrix}{\begin{bmatrix} b(x) \\ b(y) \end{bmatrix}} +
    \frac{q(x,y)}{2}\left(\dot{b}(x) + \dot{b}(y)\right) \\
    &= d\Phi_{\gamma_{id}}(b)(x,y)
\end{align*}
\noindent for ${b} \in T_{\gamma_{id}}(\Gamma_{[0,1]})$.
\end{proof}

\section{Proof of Theorem II.3}

\begin{proof}
With a bit abuse of notation, let $\Phi(\bar{\gamma}_p)(x,y) = (q_2*\bar{\gamma}_p) = q_2\big(\bar{\gamma}_p(t,{x}),\bar{\gamma}_p(t,{y})\big) \sqrt{\frac{\partial \bar{\gamma}_p(t,{x})}{\partial {x}}}\sqrt{\frac{\partial \bar{\gamma}_p(t,{y})}{\partial {y}}}$. With $t\in(-\epsilon,\epsilon)$, $\bar{\gamma}_p(t, x)$ forms a differentiable path in $\Gamma_{[0,1]}$, passing through $\gamma_{id}$ at time $t = 0$. Furthermore, let the velocity of $\bar{\gamma}_p$ at $t = 0$ be given by the tangent velocity function ${b} \in T_{\gamma_{id}}(\Gamma_{[0,1]})$. As such, the path satisfies the initial conditions:
\begin{equation*}
    \bar{\gamma}_p(0, {x}) = \gamma_{id}({x}),\; 
    \frac{\partial \bar{\gamma}_p(t, {x})}{\partial {x}}\bigg|_{t=0} = 1,\;
    \frac{\partial \bar{\gamma}_p(t, {x})}{\partial {t}}\bigg|_{t=0} = {b}({x}),\;
    \frac{\partial \bar{\gamma}_p(t, {x})}{\partial {x}\partial t}\bigg|_{t=0} = \dot{b}({x}).
\end{equation*}
We can rewrite the cost function in Equation (4) 
as
\begin{align*}
    H(f_1, (f_2*\bar{\gamma}_p)) = \|{q_1 - \Phi(\bar{\gamma}_p)}\|^2 
    = \int_{\Omega}\int_{\Omega} \left(q_1(x,y) - \Phi(\bar{\gamma}_p)(x,y)\right)^2 \; \text{d}{{x}} \text{d}{{y}}.
\end{align*}
\noindent Taking the derivative w.r.t. $t$ at $t=0$, we get, 
\begin{align*}
    \frac{d}{dt} H(\bar\gamma_p) \Big|_{t=0} &= 
    \int_{\Omega}\int_{\Omega} -2 \left(q_1 - \Phi(\gamma_{id})\right)\frac{d}{dt}\Phi(\bar{\gamma}_p)({x},{y})|_{t = 0} \; \text{d}{{x}} \text{d}{{y}} \\
    &= -2 \inner{q_1(x,y) - q_2(x,y)}{\frac{d}{dt}\Phi(\bar{\gamma}_p)({x},{y})|_{t = 0}}.
\end{align*}
\noindent By substituting in the result from {Lemma} {II.2}
, we get the desired result.
\end{proof}

\section{Proof of Lemma III.1}

\begin{proof} Let $\bar{\gamma}_p \, : \, [-\epsilon,\epsilon] \times \mathbb{S}^2 \rightarrow \mathbb{S}^2$ be a mapping $\bar{\gamma}_p(t, x)$ forming a differentiable path in $\Gamma_{\s^2}$, passing through $\gamma_{id}$ at time $t = 0$. Furthermore, let the velocity of $\bar{\gamma}_p$ at $t = 0$ be given by a tangent velocity field ${b} \in T_{\gamma_{id}}(\Gamma_{\s^2})$. As such, the path satisfies the following initial conditions:
\begin{align}
    \bar{\gamma}_p(t, {x})\bigg|_{t=0} = &\gamma_{id}({x}),\; 
    \text{det}  \frac{\partial \bar{\gamma}_p(t, {x})}{\partial {x}}\bigg|_{t=0} = |J_{\gamma_{id}}({x})| = 1,\; \nonumber \\ 
    &\frac{\partial \bar{\gamma}_p(t, {x})}{\partial {t}}\bigg|_{t=0} = {b}({x}),\;
    \frac{\partial}{\partial t}\left(\text{det}\frac{\partial\bar{\gamma}_p(t, {x})}{\partial {x}}\right)\bigg|_{t=0} = \nabla\cdot{b}({x}),
\end{align}
where $\nabla\cdot{b}$ is the divergence of the velocity vector field ${b}$. The proof follows in the same way as for Lemma {II.2}
, except for the last term in the initial conditions, which is given by 
\begin{align*}
    \frac{\partial}{\partial t}\left(\text{det}\frac{\bar{\gamma}_p(t, {x})}{\partial {x}}\right)\Big|_{t=0} &= \frac{\partial}{\partial t}\text{det} 
    \begin{bmatrix}
    \frac{\partial\bar{\gamma}_p^{(1)}}{\partial\theta} & \frac{\partial\bar{\gamma}_p^{(1)}}{\partial\phi} \\
    \frac{\partial\bar{\gamma}_p^{(2)}}{\partial\theta} &
    \frac{\partial\bar{\gamma}_p^{(2)}}{\partial\phi}
    \end{bmatrix}\Bigg|_{t=0}\\
    &= \Big(\frac{\partial\bar{\gamma}_p^{(1)}}{\partial\theta \partial t}\frac{\partial\bar{\gamma}_p^{(2)}}{\partial\phi} + \frac{\partial\bar{\gamma}_p^{(1)}}{\partial\theta}\frac{\partial\bar{\gamma}_p^{(2)}}{\partial\phi\partial t} - \frac{\partial\bar{\gamma}_p^{(2)}}{\partial\theta\partial t}\frac{\partial\bar{\gamma}_p^{(1)}}{\partial\phi} - \frac{\partial\bar{\gamma}_p^{(2)}}{\partial\theta}\frac{\partial\bar{\gamma}_p^{(1)}}{\partial\phi\partial t}\Big)\Big|_{t=0} \\
    &= \frac{\partial}{\partial\theta} \Big(\frac{\partial \bar{\gamma}_p^{(1)}}{\partial t} \Big)\Big|_{t=0} + \frac{\partial}{\partial\phi} \Big(\frac{\partial \bar{\gamma}_p^{(2)}}{\partial t} \Big)\Big|_{t=0}\\
    &= \frac{\partial}{\partial\theta}{b}^{(1)} + \frac{\partial}{\partial \phi}{b}^{(2)} \\
    &= \nabla \cdot {b}.
\end{align*}
\end{proof}

\section{Interpolation of \(f\) and \(q\) on $\mathbb{S}^2$:A Detailed Explanation}

Given that the function \(f\) is evaluated at \(V^2\) vertices on a grid, our goal is to interpolate \(f(x,y)\) for any \((x,y) \in \mathbb{S}^2 \times \mathbb{S}^2\). We achieve this using barycentric interpolation, which extends the idea of linear interpolation to simplices (e.g., triangles), making it well suited for our spherical grid.

\subsection*{1. Barycentric Interpolation in the \(x\)-Direction}

\textbf{Setup:} Consider a triangle \(\Delta x_1 x_2 x_3\) with vertices \(x_1\), \(x_2\), and \(x_3\). For a fixed \(y\) (say, \(y_1\)), we assume that the values \(f(x_1,y_1)\), \(f(x_2,y_1)\), and \(f(x_3,y_1)\) are known.

\textbf{Subtriangle Areas:} For any point \(x \in \Delta x_1 x_2 x_3\), let
\[
\begin{aligned}
A_1(x) &= \text{Area}(\Delta x\,x_2\,x_3), \\
A_2(x) &= \text{Area}(\Delta x_1\,x\,x_3), \\
A_3(x) &= \text{Area}(\Delta x_1\,x_2\,x),
\end{aligned}
\]
and denote the total area by
\[
A_x = A_1(x) + A_2(x) + A_3(x),
\]
which is the area of \(\Delta x_1 x_2 x_3\).

\textbf{Interpolation Formula:} The barycentric interpolation for the fixed \(y_1\) is then given by:
\[
f(x, y_1) = \frac{A_{1}(x)f(x_1,y_1) + A_{2}(x)f(x_2,y_1) + A_{3}(x)f(x_3,y_1)}{A_x}.
\]

\subsection*{2. Extending to Bi-variate Interpolation}

\textbf{Interpolation in \(y\):} Next, perform a similar interpolation in the \(y\)-direction. Consider a triangle \(\Delta y_1 y_2 y_3\) with vertices \(y_1\), \(y_2\), and \(y_3\). For any point \(y \in \Delta y_1 y_2 y_3\), define:
\[
\begin{aligned}
A_1(y) &= \text{Area}(\Delta y\,y_2\,y_3), \\
A_2(y) &= \text{Area}(\Delta y_1\,y\,y_3), \\
A_3(y) &= \text{Area}(\Delta y_1\,y_2\,y),
\end{aligned}
\]
with total area
\[
A_y = A_1(y) + A_2(y) + A_3(y).
\]

\textbf{Combined Interpolation:} First, interpolate in the \(x\)-direction for each fixed \(y\)-vertex:
\[
f(x,y_i) = \frac{A_{1}(x)f(x_1,y_i) + A_{2}(x)f(x_2,y_i) + A_{3}(x)f(x_3,y_i)}{A_x}, \quad i=1,2,3.
\]
Then, interpolate in the \(y\)-direction:
\[
f(x,y) = \frac{A_{1}(y)f(x,y_1) + A_{2}(y)f(x,y_2) + A_{3}(y)f(x,y_3)}{A_y}.
\]

\subsection*{3. Matrix Formulation}

The two-step interpolation can be compactly expressed in matrix form. Define the normalized area vectors:
\[
A_{xx} = \begin{bmatrix}
\frac{A_{1}(x)}{A_x} \\
\frac{A_{2}(x)}{A_x} \\
\frac{A_{3}(x)}{A_x}
\end{bmatrix}, \quad
A_{yy} = \begin{bmatrix}
\frac{A_{1}(y)}{A_y} \\
\frac{A_{2}(y)}{A_y} \\
\frac{A_{3}(y)}{A_y}
\end{bmatrix},
\]
and let \(B\) be the \(3\times 3\) matrix:
\[
B = \begin{bmatrix}
f(x_1,y_1) & f(x_2,y_1) & f(x_3,y_1) \\
f(x_1,y_2) & f(x_2,y_2) & f(x_3,y_2) \\
f(x_1,y_3) & f(x_2,y_3) & f(x_3,y_3)
\end{bmatrix}.
\]
Then the interpolation for \(f(x,y)\) can be written as:
\[
f(x,y) = A_{yy}^{\mathrm{T}} \, B \, A_{xx}.
\]

\section{Computation of \(\gamma\)'s Jacobian: Detailed Derivation.}

The warping function is defined as
$\gamma: \mathbb{S}^2 \to \mathbb{S}^2,$
where the sphere \(\mathbb{S}^2\) is parameterized by spherical coordinates \((\theta,\phi) \in [0,\pi]\times [0,2\pi)\). In our implementation, we represent \(\gamma\) in Cartesian coordinates via the mapping $\Phi$ such that
\[
\Psi(\theta,\phi)=
\begin{pmatrix}
\sin\theta\cos\phi\\[1ex]
\sin\theta\sin\phi\\[1ex]
\cos\theta
\end{pmatrix},
\]
and its inverse
\[
\Psi^{-1}(x,y,z)=
\begin{pmatrix}
\cos^{-1}(z)\\[1ex]
\tan^{-1}(y/x)
\end{pmatrix}.
\]
Thus, the warped spherical coordinate of $(\theta, \phi)$ is given by
$
(\tilde{\theta},\tilde{\phi}) = \gamma(\theta,\phi),
$ under the spherical coordinate. 

\subsection*{Tangent Space and Orthonormal Basis}
To compute the Jacobian of \(\gamma\), we differentiate along an orthonormal basis for the tangent space \(T_{\Psi(\theta,\phi)}(\mathbb{S}^2)\). A natural choice is given by
\begin{equation}
\label{eq:tangent_derivative}
\left(\frac{d\Psi}{d\theta},\; \frac{1}{\sin\theta}\frac{d\Psi}{d\phi}\right)=
\begin{pmatrix}
\cos\theta\cos\phi & -\sin\phi \\
\cos\theta\sin\phi & \cos\phi \\
-\sin\theta       & 0
\end{pmatrix}
\equiv \bigl(w^{(\theta)} \quad w^{(\phi)}\bigr).
\end{equation}

\subsection*{Numerical Approximation via Finite Differences}
For numerical stability, the partial derivatives of the warped spherical coordinates \((\tilde{\theta},\tilde{\phi})\) are approximated using a two-sided finite difference scheme. Specifically,
\begin{equation}
\label{eq:finite_difference}
\left(
\begin{matrix}
\frac{\partial\tilde{\theta}}{\partial\theta} & \frac{1}{\sin\theta}\frac{\partial\tilde{\theta}}{\partial\phi} \\[1ex]
\frac{\partial\tilde{\phi}}{\partial\theta} & \frac{1}{\sin\theta}\frac{\partial\tilde{\phi}}{\partial\phi}
\end{matrix}
\right)
\approx \frac{1}{2\delta}
\left[
\begin{matrix}
\gamma\bigl(\exp_{\Psi(\theta,\phi)}(\delta\,w^{(\theta)})\bigr) - \gamma\bigl(\exp_{\Psi(\theta,\phi)}(-\delta\,w^{(\theta)})\bigr) \\[1ex]
\gamma\bigl(\exp_{\Psi(\theta,\phi)}(\delta\,w^{(\phi)})\bigr) - \gamma\bigl(\exp_{\Psi(\theta,\phi)}(-\delta\,w^{(\phi)})\bigr)
\end{matrix}
\right]^{\!\mathrm{T}},
\end{equation}
where \(\exp_{\Psi(\theta,\phi)}(\cdot)\) denotes the exponential map on \(\mathbb{S}^2\).

In our implementation, \(\gamma\) is also represented as a vector field defined on a discrete triangular mesh, e.g., 
\(\gamma(x): \mathbb{S}^2 \to T_x(\mathbb{S}^2)\), we apply parallel transport during interpolation to ensure that the resulting vectors remain in the tangent plane. A closed-form expression for parallel transport on \(\mathbb{S}^2\) is provided in \cite{zhang2018phase}.

\subsection*{Jacobian Determinant via Change of Basis}
The Jacobian \(J_{\gamma(\theta,\phi)}\) of the diffeomorphism \(\gamma\) is an invertible linear mapping between the tangent spaces \(T_{(\theta,\phi)}(\mathbb{S}^2)\) and \(T_{(\tilde{\theta},\tilde{\phi})}(\mathbb{S}^2)\). To compute its determinant, we change the basis from the spherical coordinates \((\theta,\phi)\) to the orthonormal tangent basis \(\{w^{(\theta)},w^{(\phi)}\}\), and then back to spherical coordinates. This yields
\begin{align*}
|J_{\gamma(\theta,\phi)}| &= \left|
\begin{pmatrix}
1 & 0 \\
0 & \sin\tilde{\theta}
\end{pmatrix}
\begin{pmatrix}
\frac{\partial\tilde{\theta}}{\partial\theta} & \frac{\partial\tilde{\theta}}{\partial\phi} \\
\frac{\partial\tilde{\phi}}{\partial\theta} & \frac{\partial\tilde{\phi}}{\partial\phi}
\end{pmatrix}
\begin{pmatrix}
1 & 0 \\
0 & \frac{1}{\sin\theta}
\end{pmatrix}
\right| \\
&= \sin\tilde{\theta}\,
\left|\begin{matrix}
\frac{\partial\tilde{\theta}}{\partial\theta} & \frac{1}{\sin\theta}\frac{\partial\tilde{\theta}}{\partial\phi} \\[1ex]
\frac{\partial\tilde{\phi}}{\partial\theta} & \frac{1}{\sin\theta}\frac{\partial\tilde{\phi}}{\partial\phi}
\end{matrix}\right|.
\end{align*}
Here, the left and right matrices perform the necessary change of basis to and from the orthonormal tangent space. Importantly, by computing derivatives in the directions \(w^{(\theta)}\) and \(w^{(\phi)}\) rather than directly in \((\theta,\phi)\), we avoid singularities when \(\sin\theta = 0\).

\section{Spatial Derivative of \(q: \mathbb{S}^2 \times \mathbb{S}^2 \rightarrow \mathbb{R}_{\ge 0}\)}
To compute the directional derivative of \(q(x,y): \mathbb{S}^2 \times \mathbb{S}^2 \to \mathbb{R}_{\ge 0}\) along the vector field \(\begin{bmatrix} b(x) \\ b(y) \end{bmatrix}\), as required in Lemma~III.1, we use a finite difference approach. Let \(\{w_x^{(1)},w_x^{(2)}\}\) be an orthonormal basis for the tangent space \(T_x(\mathbb{S}^2)\). Any vector \(b(x)\) can then be written as
\[
b(x) = a_x^{(1)}\, w_x^{(1)} + a_x^{(2)}\, w_x^{(2)},
\]
with coefficients \(a_x^{(1)}\) and \(a_x^{(2)}\).

Under this basis, the partial derivative of \(q(x,y)\) with respect to \(x\) can be represented as
\[
\frac{\partial q(x,y)}{\partial x} = \begin{bmatrix}
dq_{x}\bigl(w_x^{(1)},y\bigr) & dq_{x}\bigl(w_x^{(2)},y\bigr)
\end{bmatrix},
\]
where each directional derivative is defined by
\[
dq_{x}  \bigl(w_x^{(i)},y\bigr) \coloneqq \lim_{\epsilon \rightarrow 0} \frac{1}{\epsilon}\left(q\bigl(\exp_x(\epsilon\,w_x^{(i)}),y\bigr)-q(x,y)\right).
\]
Here, \(\exp_x\) denotes the exponential map at \(x\) on \(\mathbb{S}^2\), which maps a tangent vector to a point on the sphere. In practice, we approximate \(q\bigl(\exp_x(\epsilon\,w_x^{(i)}),y\bigr)\) by choosing a small \(\epsilon\) and performing barycentric interpolation, as described previously.

Similarly, the partial derivative with respect to \(y\) can be computed using an orthonormal basis \(\{w_y^{(1)},w_y^{(2)}\}\) for \(T_y(\mathbb{S}^2)\), with the corresponding expression
\[
\frac{\partial q(x,y)}{\partial y} = \begin{bmatrix}
dq_{y}\bigl(x,w_y^{(1)}\bigr) & dq_{y}\bigl(x,w_y^{(2)}\bigr)
\end{bmatrix},
\]
and coefficients \(b(y)=a_y^{(1)}\,w_y^{(1)}+a_y^{(2)}\,w_y^{(2)}\).

Thus, the total directional derivative along the vector field \(\begin{bmatrix} b(x) \\ b(y) \end{bmatrix}\) is given by $
dq_{x}\bigl(w_x^{(1)},y\bigr)a_x^{(1)} + dq_{x}\bigl(w_x^{(2)},y\bigr)a_x^{(2)} + dq_{y}\bigl(x,w_y^{(1)}\bigr)a_y^{(1)} + dq_{y}\bigl(x,w_y^{(2)}\bigr)a_y^{(2)}.$

\section{Extension to \(\Omega = \mathbb{S}^2_1 \cup \mathbb{S}^2_2\)}
In our application, the domain of interest for cortical surface registration comprises two hemispheres. Therefore, the methods developed above must be extended to work on the domain
\[
\Omega = \Omega_1 \cup \Omega_2,
\]
where \(\Omega_1 \cap \Omega_2 = \emptyset\), with \(\Omega_1 = \s^2 \) representing the left hemisphere and \(\Omega_2 = \s^2 \) representing the right hemisphere. In this setting, the diffeomorphism \(\gamma\) is defined piecewise as
\[
\gamma(x) =
\begin{cases}
\gamma^1(x) & \text{if } x \in \Omega_1, \\
\gamma^2(x) & \text{if } x \in \Omega_2.
\end{cases}
\]

We then redefine the mapping \(\Phi\) from Eq.~(5) 
as follows:
\begin{equation}\label{eq:full_phi}
\Phi(\gamma)(x,y) =
\begin{cases}
q\bigl(\gamma^1(x),\gamma^1(y)\bigr)\sqrt{J_{\gamma^1}(x)}\sqrt{J_{\gamma^1}(y)}, & \text{if } x \in \Omega_1,\; y \in \Omega_1, \\[1ex]
q\bigl(\gamma^2(x),\gamma^1(y)\bigr)\sqrt{J_{\gamma^2}(x)}\sqrt{J_{\gamma^1}(y)}, & \text{if } x \in \Omega_2,\; y \in \Omega_1, \\[1ex]
q\bigl(\gamma^1(x),\gamma^2(y)\bigr)\sqrt{J_{\gamma^1}(x)}\sqrt{J_{\gamma^2}(y)}, & \text{if } x \in \Omega_1,\; y \in \Omega_2, \\[1ex]
q\bigl(\gamma^2(x),\gamma^2(y)\bigr)\sqrt{J_{\gamma^2}(x)}\sqrt{J_{\gamma^2}(y)}, & \text{if } x \in \Omega_2,\; y \in \Omega_2.
\end{cases}
\end{equation}

Assuming that \(q(x,y)\) is continuous and differentiable even when \(x\) and \(y\) lie in different subdomains, each of the cases in Eq.~\eqref{eq:full_phi} can be treated as before. Consequently, the underlying theory remains unchanged. In practice, when determining the optimal warping function to align with the ConCon functions, we propose an iterative procedure in which the warp for each hemisphere is updated individually.

Note that the first and last cases in Eq.~\eqref{eq:full_phi} depend independently on \(\gamma^1\) and \(\gamma^2\), whereas the middle two cases are symmetric. Therefore, we can define the differential of \(\Phi\) with respect to each warping function separately.

\begin{lemma}\label{dphi_cortical}
Let the domain $\Omega = \Omega_1 \cup \Omega_2$, and $\gamma \in \Gamma_\Omega$.   The differential of \(\Phi : \Gamma_\Omega \to [q]\), while holding \(\gamma^2\) as fixed and given and evaluated at the identity warp \(\gamma^1_{id}\), is given by
\begin{equation}
\label{eq:dphi_cortical}
\bigl(d\Phi_{\gamma^1_{id}}(b)\bigr)(x,y) =
\begin{cases}
\frac{\partial q(x,y)}{\partial x}\, b(x) + \frac{\partial q(x,y)}{\partial y}\, b(y) + \frac{q(x,y)}{2}\Bigl( \nabla \cdot b(x) + \nabla \cdot b(y) \Bigr), & \text{if } x, y \in \Omega_1, \\[1ex]
\frac{\partial q(x,y)}{\partial x}\, b(x) + \frac{q(x,y)}{2}\,\nabla \cdot b(x), & \text{if } x \in \Omega_1,\; y \in \Omega_2, \\[1ex]
\frac{\partial q(x,y)}{\partial y}\, b(y) + \frac{q(x,y)}{2}\,\nabla \cdot b(y), & \text{if } x \in \Omega_2,\; y \in \Omega_1, \\[1ex]
0, & \text{otherwise.}
\end{cases}
\end{equation}
Here, \(\nabla \cdot b\) denotes the divergence of the vector field \(b \in T_{\gamma^1_{id}}(\Gamma_{\s^2})\). The differential of \(\Phi\) while holding \(\gamma^1\) fixed is defined analogously.
\end{lemma}

\begin{proof}
Let 
\[
\bar{\gamma}_p : [-\epsilon,\epsilon] \times \bigl(\Omega \bigr) \to \Omega
\]
be a mapping that defines a differentiable path in \(\Gamma_\Omega\) passing through \(\gamma^1_{id}\) at \(t = 0\) with $\gamma^2$ fixed; that is, \(\bar{\gamma}_p(0,u) = \gamma^1_{id}(u)\) for $u \in \Omega_1$. The proof for the first case (when \(x,y \in \Omega_1\)) follows similarly to Lemmas~II.2 
and III.1
.

For the second and third cases, let \(u\) be any point in $\Omega$. If we fix \(\gamma^2\) to the identity, then for any \(u \in \Omega_2\) the velocity of \(\bar{\gamma}_p\) at \(t = 0\) is zero. Hence, for \(u \in \Omega_2\) the path satisfies the following initial conditions:
\begin{align}\label{initialconditions}
\bar{\gamma}_p(t,u)\big|_{t=0} &= \gamma_{id}^2(u), \\
\det\frac{\partial \bar{\gamma}_p(t,u)}{\partial u}\Big|_{t=0} &= |J_{\gamma^2_{id}}(u)| = 1, \nonumber \\ 
\frac{\partial \bar{\gamma}_p(t,u)}{\partial t}\Big|_{t=0} &= 0, \quad
\frac{\partial}{\partial t}\left(\det\frac{\partial \bar{\gamma}_p(t,u)}{\partial u}\right)\Big|_{t=0} = 0. \nonumber
\end{align}
Since the velocity of \(\bar{\gamma}_p\) at \(t = 0\) is given by \(b \in T_{\gamma^1_{id}}(\Gamma_{\s^2})\) when \(u \in \Omega_1\), the result can be derived easily as in the previous case.
\end{proof}

Combining these results, we can now define the continuous connectivity registration algorithm by employing the derivative of the cost function with respect to the warp function for each hemisphere separately. The directional derivative of the cost function \(H\) with respect to \(\gamma^1\) in the direction \(b_i \in T_{\gamma^1_{id}}(\Gamma_{\s^2})\) is given by

\begin{align}
\label{eq:optomega}
\nabla_{b_i} H_{\gamma^1} &= \int_{x \in \Omega_1} \int_{y \in \Omega_1} \Bigl\{ q_1(x,y) - q_2(x,y) \Bigr\} \notag \\
&\quad \times \left\{ \begin{bmatrix}\frac{\partial q_2(x,y)}{\partial x} & \frac{\partial q_2(x,y)}{\partial y}\end{bmatrix} \begin{bmatrix} b_i(x) \\ b_i(y) \end{bmatrix} + \frac{q_2(x,y)}{2} \Bigl( \nabla \cdot b_i(x) + \nabla \cdot b_i(y) \Bigr) \right\} dx\,dy \notag \\
&\quad + \int_{x \in \Omega_1} \int_{v \in \Omega_2} \Bigl\{ q_1(x,v) - q_2(x,v) \Bigr\} \notag \\
&\quad \times \left\{ \frac{\partial q_2(x,v)}{\partial x}\, b_i(x) + q_2(x,v)\,\nabla \cdot b_i(x) \right\} dx\,dv.
\end{align}

\section{Algorithms for Defining a Template for Population-based Registration}

\subsubsection{Center of $\{[q_j]\}$}

To find a template, the intuitive approach would be to take the cross-sectional mean of the functions $f_j \in \mathcal{F}$. However, this would be a poor estimator of the mean because we do not have a distance in $\mathcal{C}$ that is invariant to warping. Instead, we can take the mean of the orbits $\{[q_j]\}$ which live in the the quotient space $\mathcal{S} = \s^\infty / \Gamma$. We have already defined a distance between elements $q_1,q_2 \in \s^\infty$ in section II-A
. Since the group \(\Gamma\) acts isometrically on \(\s^\infty\), under appropriate regularity conditions the distance defined on \(\s^\infty\) descends via the quotient map to induce a well-defined Riemannian structure on the quotient space \(\mathcal{S}\).
 As such, we can define the Karcher mean \cite{RN82} of points in $\mathcal{S}$ as
\begin{equation}\label{eq:frechetmean}
    [\mu] = \argmin_{[q] \in \mathcal{S}} \sum_{j=1}^N d([q],[q_j])^2,
\end{equation}
\noindent where $d([q],[q_j]) = \min _{\gamma_1,\gamma_2 \in \Gamma}\cos^{-1}(\inner{(q*\gamma_1)}{(q_j*\gamma_2)})$. 
Unfortunately, there is no analytic form to solve \eqref{eq:frechetmean}, and it must instead be solved using an iterative algorithm \cite{RN1,RN80} presented in Algorithm \ref{alg:frechet}. 

\begin{algorithm}
\caption{Karcher mean of $\{[q_j]\}$}\label{alg:frechet}
\begin{algorithmic}
\State Initialise $\mu$ as one of $q_j$'s. 
\State Set $k = 0$, and $\epsilon_1, \epsilon_2$ to be small positive values
\vspace{0.2em}
\While{$||\bar{v}|| > \epsilon_1$}
\State Align each $q_j$ to $\mu_k$, and denote them as $\tilde{q}_j$
\vspace{0.2em}
\State For each $j = 1,2,\dots,N$ compute $v_i \gets \log_{\mu_k}(\tilde{q}_i)$
\vspace{0.2em}
\State \text{Compute the average direction $\bar{v} \gets \tfrac{1}{N}\sum_{i=1}^N v_i$}
\vspace{0.2em}
\State \text{Update } $\mu_{k+1} \gets \exp_{\mu_k}(\epsilon_2 \bar{v})$ 
\vspace{0.2em}
\State $k \gets k + 1$
\EndWhile
\end{algorithmic}
\vspace{0.2em}
\textit{\noindent The final orbit $[\mu_{k}]$ is considered as $[\mu]$, and the exponential and logarithmic maps are defined on $\s^\infty$.}
\end{algorithm}

 The mean estimate can be susceptible to outliers, as such, it may be desirable to take the median for the template instead. An algorithm for finding the median is very similar to the gradient-based approach to finding the mean \cite{RN79}. After getting the template orbit, we will want to select an element $\mu \in [\mu]$ that is in some way optimal.

\subsubsection{Center of the Orbit $[\mu]$}

Following on from \cite{RN1,RN26}, we assume that the mean of the warping functions needed to align the connectivity functions in a sample should be (close to) the identity, i.e. $\mathbb{E}\{\gamma(x)\} = x$ for $x \in \Omega$. So then, we find the template $\mu^*$ from $[\mu]$ as follows. 

\begin{definition}
For a given set of functions $q_1,q_2,...,q_N \in \mathcal{Q}$, and orbit, $[\mu] \in \mathcal{S}$, define the element $\mu^*$ as the center of $[\mu]$ with respect to the set $\{q_j\}$ if the warping functions $\gamma_j \in \Gamma$, where $\gamma_j = \argmin_{\gamma \in \Gamma} \|\mu^* - (q_j * \gamma)\|^2$, have the Karcher mean $\gamma_{id}$. 
\end{definition}

\noindent The algorithm to find the center of the orbit is presented in \cite{RN1}, but we will state it here in a more general form.

\begin{algorithm}
\caption{The center of a orbit $[\mu]$ w.r.t. the set $\{q_j\}$}\label{alg:template}
\begin{enumerate}
    \item For each $q_j \in \mathcal{Q}$, find the warping functions $\gamma_j \in \Gamma$ by solving $\gamma_j = \argmin_{\gamma \in \Gamma} \|\mu - (q_j * \gamma)\|^2$ using Algorithm 2 in the main text.
    \item Compute the Karcher mean $\bar{\gamma}_p$ of all $\gamma_j$ using Algorithm \eqref{alg:frechet}, where the exponential and logarithmic maps are on $\Gamma$ instead of $\mathcal{S}$, and we initialise the algorithm using $\gamma_{id}$. 
    \item The center of $[\mu]$ w.r.t. $\{q_j\}$ is given by $\mu^* = (\mu_k * \bar{\gamma}_p^{-1})$, where $\mu_k$ is the output from Algorithm 2 in the main text.
\end{enumerate}
\end{algorithm}

\subsubsection{A complete registration algorithm for connectivity functions}
Now we have all the elements we need for a complete registration algorithm for connectivity functions, which we present in Algorithm \eqref{alg:registration}. 

\begin{algorithm}
\caption{Registration of ConCon Functions}\label{alg:registration}
\begin{enumerate}
    \item Given functions $f_1,f_2,\dots,f_N \in \mathcal{F}$, compute $q_1,q_2,\dots,q_N \in \mathcal{Q}$ using the $Q$ map (Equation (3) in the main text).
    \item Compute the Karcher mean (or median) orbit $[\mu]$ of $[q_1],[q_2],\dots,[q_N] \in \mathcal{S}$ using Algorithm \eqref{alg:frechet}.
    \item Find the center of $[\mu]$ w.r.t. $\{[q_j]\}$ using Algorithm \eqref{alg:template}; call it $\mu^*$.
    \item For each $q_j \in \mathcal{Q}$, find the warping functions $\gamma_j^* \in \Gamma$ by solving $\gamma_j^* = \argmin_{\gamma \in \Gamma} \|\mu^* - (q_j * \gamma)\|^2$ using Algorithm 2 
    in the main text.
    \item Compute the aligned connectivity functions $f^*_j = (f_j * \gamma_j^*)$.
    \item Return the template $\mu^*$, the warping functions $\{\gamma_j^*\}$, and the aligned functions $\{f_j^*\}$. 
\end{enumerate}
\end{algorithm}

\section{More Results for Simulation Study}

\begin{figure}[th!]
     \centering
    \includegraphics[width=\columnwidth]{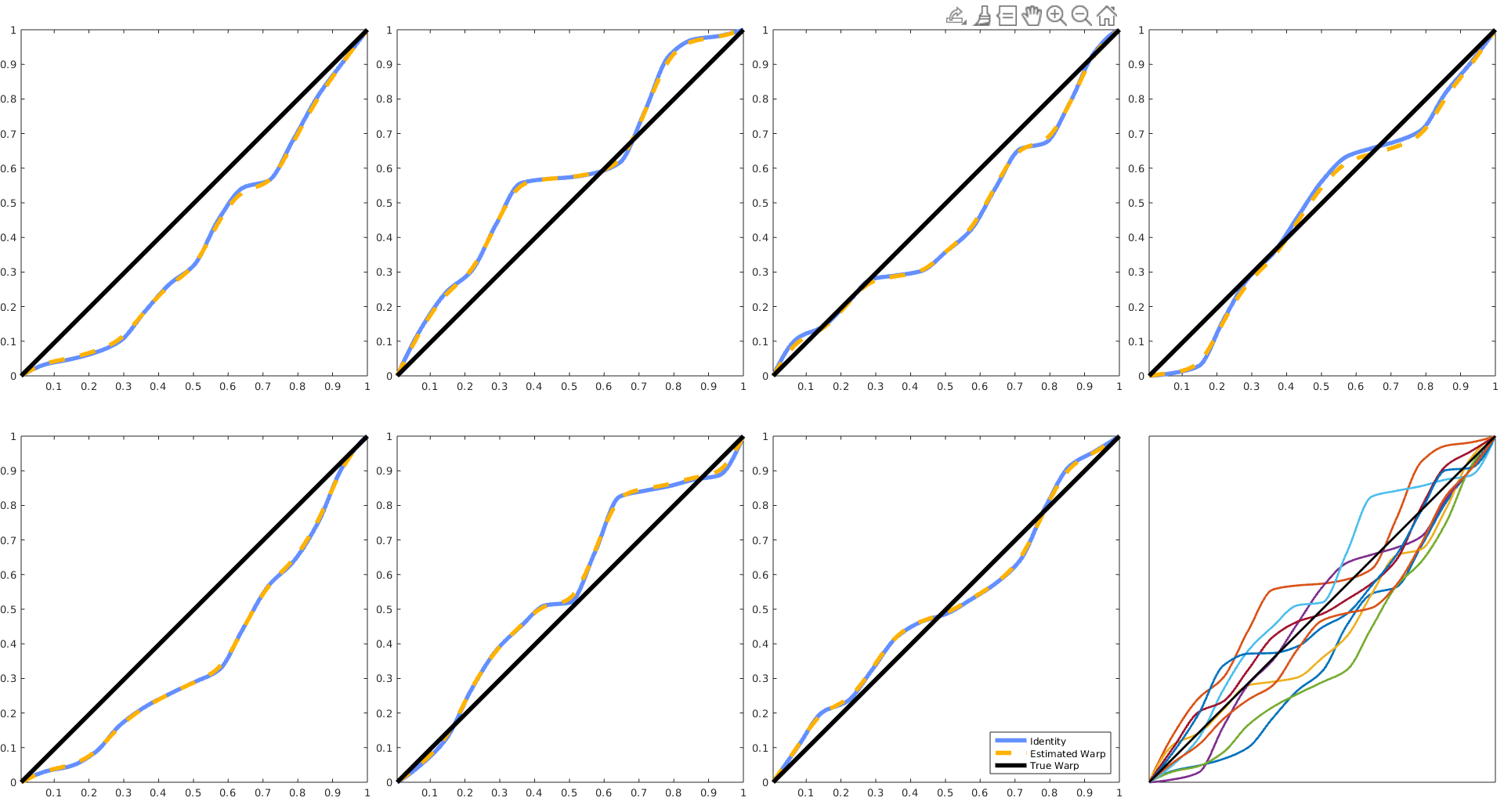}
    \caption{Shows 7 randomly selected warping functions from the simulation and their estimates, demonstrating that ENCORE works well regardless of the shape of the warping function.}
    \label{fig:simulation_warps}
\end{figure}

\begin{table}[th!]
\centering
\begin{tabular}{|c|c|c|c|}
\hline
N & $\mu_1$ & $\mu_2$ & $\sigma^2$ \\
\hline
2500 & 0.1 & 0.9 & 0.5 \\ 
2500 & 0.1 & 0.3 & 0.5 \\ 
2500 & 0.5 & 0.6 & 0.4 \\ 
2500 & 0.2 & 0.6 & 0.2 \\ 
8000 & 0.55 & 0.95 & 0.3 \\ 
2000 & 0.4 & 0.8 & 0.6 \\ 
\hline
\end{tabular}
\caption{For simulated ConCon functions, we generated six sets of endpoints, each set containing N endpoint pairs from normal distributions with means $\mu_1$, $\mu_2$, and variance $\sigma^2$.}
\end{table}
\clearpage

\section{More Results for Real Data Analysis}

\begin{figure}[th!]
     \centering
    \includegraphics[width=\columnwidth]{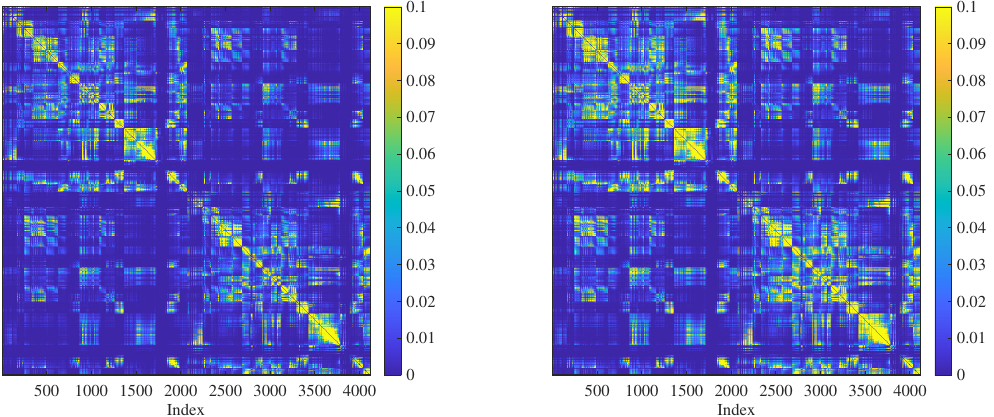}
    \vspace{2mm} 
    \parbox{\columnwidth}{
        \centering
        (A) \hspace{0.45\columnwidth} (B)
    }
    \caption{Shows the templates generated from different groups of selected subjects.}
    \label{fig:template}
\end{figure}

\begin{figure}[th!]
     \centering
    \includegraphics[width=\columnwidth]{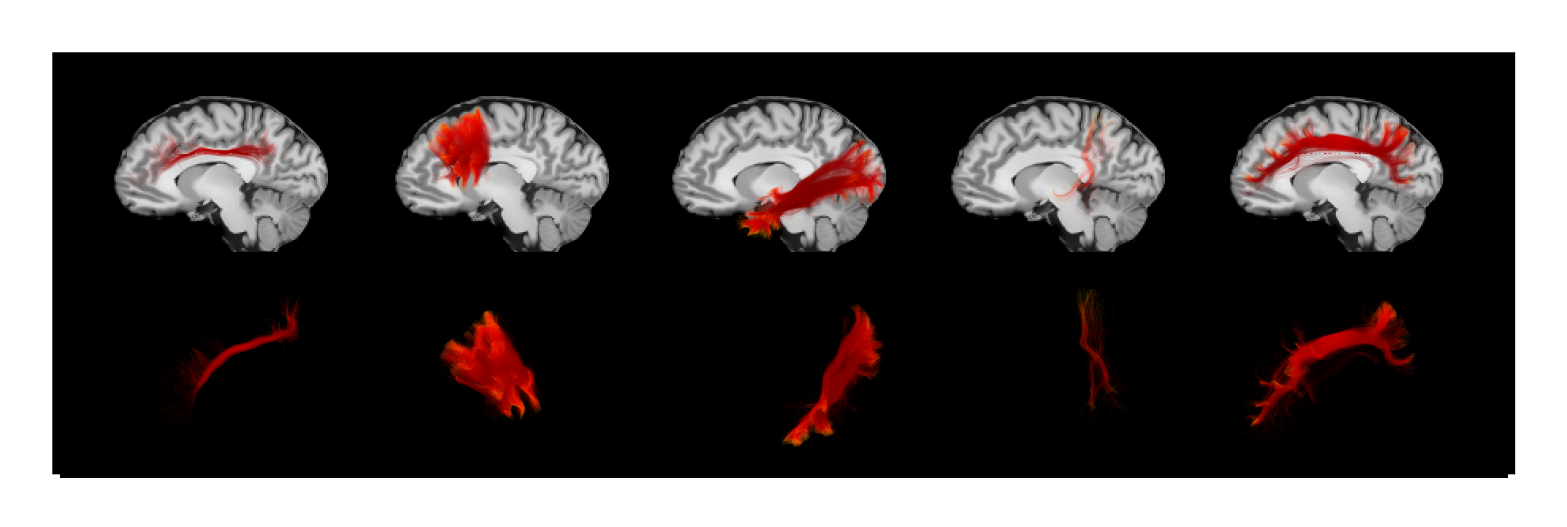}
    \caption{Shows the five major fiber bundles: bilateral Middle Longitudinal Fasciculus (MdLF), Inferior and Superior Longitudinal Fasciculus (ILF, SLF), Frontal Aslant Tract (FAT), and Anterior Cingulate Cortex (CG\_An) used in the DICE analysis.}
    \label{fig:bundles}
\end{figure}

\bibliography{references}